\documentclass{aa}
\usepackage{graphicx}
\usepackage{natbib}
\usepackage{txfonts}
\usepackage{epstopdf}
\usepackage[utf8]{inputenc}
\usepackage[english]{babel}
 
\bibpunct{(}{)}{;}{a}{}{,} 

\def\apjs{ApJS}
  
\def\jgr{J. Geophys. Res.}
   
\def\grl{Geophys. Res. Lett.}

\def\aap{Astron. Astrophys.}

\def\apj{Astrophys. J.}

\def\solphys{Sol. Phys.}

\def\mnras{Mon. Not. R. Astron. Soc.}

\begin{document}

\title{Inflection point in the power spectrum of stellar brightness variations. I. The model}
\author{A.I. Shapiro \inst{1} \and E.M. Amazo-G\'{o}mez \inst{1,2} \and N.A. Krivova \inst{1} \and S.K. Solanki \inst{1,3}}
\offprints{A.I. Shapiro}

\institute{Max-Planck-Institut fur Sonnensystemforschung, Goettingen, Germany\\
\email{shapiroa@mps.mpg.de}
\and Georg-August Universit\"{a}t G\"{o}ttingen, Institut f\"{u}r Astrophysik, Friedrich-Hund-Platz 1, 37077 G\"{o}ttingen, Germany
\and School of Space Research, Kyung Hee University, Yongin, Gyeonggi 446-701, Korea}

\date{Received ; accepted }

\abstract 
{
Considerable effort has been put into using light curves observed by space telescopes such as CoRoT, Kepler and TESS for determining stellar rotation periods. While rotation periods of active stars can be reliably determined, the light curves of many older and less active stars (e.g. stars similar to the Sun) are quite irregular, which hampers determination of their periods.}
{We examine the factors causing the irregularities in stellar brightness variations and develop a method for determining rotation periods of low activity stars with irregular light curves.}
{We extend the Spectral And Total Irradiance Reconstruction (SATIRE) approach for modelling solar brightness variations to Sun-like stars. We calculate the power spectra of stellar brightness variations for various combinations of parameters defining the surface configuration and evolution of stellar magnetic features.} 
{The short lifetime of spots in comparison to the stellar rotation period  as well as the interplay between spot and facular contributions to brightness variations of stars with near solar activity cause irregularities in their light curves. The power spectra of such stars often lack a peak associated with the rotation period. Nevertheless, the rotation period can still be determined by measuring the period where the concavity of the power spectrum plotted in the log-log scale changes sign, i.e. by identifying the position of the inflection point.}
{The inflection point of the (log-log) power spectrum is found to be a new diagnostic for stellar  rotation periods that is shown to work even in cases where the power spectrum shows no peak at the rotation rate.}
\titlerunning{Decrypting brightness variations of Sun-like stars}

\maketitle
\section{Introduction}\label{sect:intro}
The magnetic features on stellar surfaces lead to quasi-periodic variations in stellar brightness as stars rotate. While such rotation variations were first detected with ground-based instrumentation \citep[see, e.g.][]{radicketal1998}, most of the data have been accumulated with planet-hunting spaceborne missions aimed at detecting planetary transits via photometric monitoring. In particular, CoRoT \citep{COROT2, COROT} and Kepler \citep{KEPLER} telescopes provided photometric time series for several hundred thousand stars. Even more data is expected from the recently launched TESS mission \citep{TESS} and the future PLATO mission \citep{PLATO}.

The interest in studying stellar brightness variations is twofold. First, they provide information on the stars themselves, e.g. their rotation periods or their magnetic cycles. Second, a quantitative assessment of stellar variability is needed for better detection and characterization of extra-solar planets.

Of particular interest are studies of stellar rotation periods. Stellar rotation is closely linked to stellar magnetic activity and age \citep{Skumanich1972}. Consequently, surveys of stellar rotation periods are the basis for calibrating  gyrochronology relationships between rotation period, color, and age \citep[cf.][]{McQuillan2013}, for understanding the Galactic star formation history \citep[cf.][]{Davenport2017,Davenport2018}, and for constraining properties of the magnetic braking \citep{Metcalfe2016}. 

The light curves of many, especially young and active stars, look almost like a sine wave \citep[see, e.g. Fig.~4 from][]{Timo2013}. The rotation period of such stars manifests itself as a clear peak in the Lomb-Scargle periodogram \citep{LS_general} or a series of equidistant peaks in the autocorrelation function \citep{ACF_general} of their light curves. Consequently, in the numerous studies aimed at determining stellar rotation periods employing Kepler and CoRoT data \citep{Walkowicz_Basri2013, Timo2013,McQuillan2013,Garcia2014,Buzasi2016,Angus2018} most of the obtained rotation periods are of such stars.

The largest available surveys of stellar rotation periods have been compiled by \cite{Timo2013} using Lomb-Scargle periodograms and by \cite{McQuillan2013} using autocorrelation analysis. They determined rotation periods in, respectively, 24124 and 34030 presumably main sequence Kepler stars. Another approach was taken by \cite{Garcia2014}, who concentrated on Kepler stars with measured pulsations and, in addition, to the autocorrelation analysis employed wavelet power spectra. They determined rotation periods in 310 out of 540 considered targets. 

As successful as they are, the aforementioned approaches are based on the assumption that stellar light curves  have a regular temporal profile. This is a valid assumption for young and active stars but it fails for many old and less active stars. For such stars the complex configuration of magnetic features and their relatively rapid evolution lead to rather complex light curves and render the period determination very difficult.

The most prominent example of such a star with complex light curve is our Sun. Solar  short-term variability (i.e. variability on timescales of up to a few solar rotation periods) has a highly irregular temporal profile \citep[see, e.g. Fig.~1 from][]{Shapiroetal2016}. The main reason for this is that only very few sunspots last longer than the solar rotation period so that the short-term variability of solar brightness is strongly affected by sunspot evolution. 
Furthermore, \cite{Sasha_NAT} showed that the global wavelet power spectrum of solar brightness variations, calculated over the period 1996-2015, does not have a clear rotation peak due to the compensation of facular and spot contributions to solar brightness variability. They showed that since faculae have much longer lifetimes than spots (e.g. facular features can easily last for a few solar rotations), their contribution to solar brightness variability has a very pronounced peak at the solar rotation period.
The peak in the spot component of solar brightness variations is much less pronounced 
but the spot component is stronger on timescales around the solar rotation period (i.e. at about 10--50 days). As a result, two peaks almost fully cancel each other. This is in agreement with other studies \citep[see, e.g.][]{Lanza_Shkolnik2014,Aigrain2015} that found that the true rotation period of the Sun would not be detectable at intermediate and high levels of solar activity, when the spot contribution to solar brightness variations wipes out the rotation peak in the facular component. At the same time, the solar rotation period is easily detectable during activity minimum when the brightness variations are brought about by long-lived faculae.


A good understanding of the physical phenomena determining solar variability might be helpful for solving problems posed by stellar data. For example, 
\cite{Timo2018} suggested that the dearth of the intermediate stellar rotation periods in the Kepler sample \citep[see, e.g.][]{McQuillan2013,Davenport2017,Davenport2018} can be partially caused by the compensation of facular and spot contributions to brightness variability (similar to the solar case) and the consequent inability to detect rotation periods of such stars.
Therefore the dearth in the {\it observed} period distribution does not necessarily implies an under-representation in the real period distribution.

We suggest that the irregularity of stellar light curves is an important factor in explaining why rotation periods cannot be determined for the majority of stars in the Kepler field (e.g. the success rate of \cite{McQuillan2013} is only 25.5\% since they applied the autocorrelation method to 133030 stars). Recently, \cite{vanSaders2018} showed that the success rate of period determinations strongly decreases with increasing stellar effective temperature and such a decrease cannot be explained by the simultaneous decrease of the amplitude of stellar brightness variations. This is in line with our suggestion, since spot lifetimes are expected to decrease with stellar effective temperature \citep{Giles2017} and, consequently, the period determination gets more difficult.



In this paper we employ an approach similar to that taken by the SATIRE model \citep[which stands for Spectral And Total Irradiance Reconstruction,][]{fliggeetal2000,krivovaetal2003}, originally developed for modeling solar brightness variations, to synthesize stellar light curves and their power spectra. We do this as a function of lifetimes of spots and faculae, the ratio between facular and spot stellar surface-area coverage, and stellar inclination (i.e. the angle between the direction to the observer and stellar rotation axis). We specify the conditions under which the rotation peak in stellar power spectra disappears and show that even in such cases the rotation period can still be determined from the high-frequency tail of the power spectrum. In particular, we
calculate the frequency where the concavity of the power spectrum plotted on the log-log scale changes sign (in other words the steepest point of the power spectrum). Such a frequency corresponds to the inflection point in the power spectrum of stellar brightness variations. We demonstrate that the position of the inflection point is proportional to the stellar rotation frequency and can be used as a proxy for its determination. All in all, we show that the power spectrum of stellar brightness variations is a sensitive tool for studying stellar rotation and magnetic activity. 

The SATIRE approach has been extensively validated against various solar data \citep[see e.g.][and references therein]{balletal2014, yeoetal2014}. Recently SATIRE was used to show that observed solar brightness variations can be explained with remarkable accuracy by the joint action of only two sources, the surface magnetic field and granular convection \citep{Sasha_NAT}. This result puts us in a strong position for modeling  {brightness} variations of Sun-like stars.

This paper is restricted to modeling for validating the proposed approach, while its application to available stellar data is the subject of forthcoming papers. The rest of the paper is structured as follows: in Sect.~\ref{sect:model} we describe the model used to synthesize the light curves presented in this study. In Sect.~\ref{sect:spots} we consider an illustrative case of stars whose variability is exclusively brought about by dark spots. A more realistic case of stars with dark spots and bright faculae is detailed in Sect.~\ref{sect:fac}. The impact of various properties of stellar magnetic features on the position of the inflection point is outlined in Sect.~\ref{sect:rel}, while the dependence of the inflection point position on the level of stellar magnetic activity is presented in Sect.~\ref{subsect:activity}. Finally, conclusions are drawn in Sect.~\ref{sect:conc}.

\section{Model description}\label{sect:model}
Strong concentrations of magnetic field emerging on the stellar surface lead to the formation of active regions, encompassing magnetic features such as dark spots and bright faculae \citep[see, e.g., review by][]{Sami_B}. The transits of these regions over the visible stellar disk as the star rotates as well as their evolution are dominant sources of brightness variations in Sun-like stars on timescales from about a day.

In our model we construct active regions as a mixture of spot and facular areas.  We note that a model based on such an assumption would not be suitable for calculating stellar brightness variations on the timescale of the magnetic activity cycle since it does not account for the emergence of the ephemeral active regions \citep[see, e.g.][and references therein]{Dasi2016}. At the same time the variability on the timescale of stellar rotation is brought about by the largest facular and spot features, which usually emerge together. Therefore such a simple model of stellar active regions is expected to be appropriate for modeling stellar rotation variability and is often employed in the literature \citep[see, e.g.][]{Lanza2003,Lanza2009,Gondoin2009,Borgniet2015,Morris2018}.

The size of active regions is assumed to be much smaller than the stellar radius, which is a good assumption for the Sun and for stars with near-solar levels of magnetic activity. Consequently, we did not consider the exact geometrical shape of active region and its spot and facular components, prescribing the same value of the foreshortening factor for the entire region when computing its visible solid angle. 


At each moment of time $t$ the stellar brightness at the wavelength $\lambda$ is given by
\begin{equation}\label{eq:F}
F(\lambda,t)=F_{Q} (\lambda) + \sum_i \Omega_i (t) \, \cdot \left ( \phi_{i,F}(t) \, C_F \left (\lambda, \vec{r_i} \right) + \phi_{i,S}(t) \, C_S \left (\lambda, \vec{r_i} \right) \right), 
\end{equation}
where $F_Q(\lambda)$ is the brightness of a quiet star, i.e. a star without any active regions. The summation is performed over all active regions visible at time $t$ and $\Omega_i(t)$ is a solid angle of the $i$-th active region seen from the vantage point of observer. By changing the number  of regions on the stellar surface we could simulate stars with different magnetic activity. Factors $\phi_{i,F}(t)$ and $\phi_{i,S}(t)$ are fractions of spot and facular parts of the area  of $i$-th active region, respectively (with $\phi_{i,F}(t)+\phi_{i,S}(t)=1$). $C_F(\lambda, \vec{r_i})$ and $C_S(\lambda, \vec{r_i})$ are spectral contrasts of faculae and spots relative to the quiet stellar regions along the direction to the $i$-th active region $\vec{r_i}$.  

For simplicity, we consider here only stars rotating as a solid body. We assume that the emergence of stellar active regions follows the solar latitudinal distribution and happens in the activity belts, i.e. in between the latitudes of 5$^{\circ}$ and 30$^{\circ}$ \citep[see, e.g.][]{knaacketal2001}. This should be a good approximation for slowly rotating stars like the Sun \citep{polar_spot1}, which are the main focus of our study. 

As will be shown below the growth phase of the active regions does not have a strong effect on our results so that it is neglected in most of the experiments. In other words, we start tracking the region and its effect on stellar brightness only after it reaches its maximum area.

We have utilized spectra of the quiet Sun, faculae, spot umbra, and spot penumbra calculated by \cite{Unruhetal1999} with the ATLAS9 radiative transfer code \citep{kurucz1992,ATLAS9_CK}. Following \cite{wenzleretl2006} and \cite{balletal2012} we compute sunspot spectra as a mixture of 80\% penumbral and 20\% umbral spectra. 

 {The \cite{Unruhetal1999} spectra have been proved to be reliable for modeling solar brightness variations \citep[see e.g. reviews by][and references therein]{TOSCA2013, MPS_AA} and, consequently, we expect them to be applicable to modeling stars with near-solar fundamental parameters. At the same time,  the profile of the high-frequency tail of the power spectrum and, consequently, the position of the inflection point depends on the the centre-to-limb variations of  brightness contrasts of stellar magnetic features. They, in turn, depend on the fundamental stellar parameters.  The apparatus for calculating contrasts of magnetic features in stars with various fundamental parameters is becoming available \citep[see][]{norrisetal2017,witzkeetal2018,Salhabetal2018} so that we plan to generalize our study in one of the forthcoming publications. Presently available simulations of facular contrasts at stars with different effective temperatures \citep[see, e.g., Fig.~5.16 from][]{norris_PhD} 
indicate that results presented in this study are applicable to at least stars from late F to early K spectral types.}

All the light curves presented in this study are calculated as they would be seen by the Kepler telescope, i.e. by multiplying Eq.~\ref{eq:F} with the Kepler total spectral efficiency and integrating it over all relevant wavelengths. The simulations are performed with 6-hour cadence. We have checked that the decrease of the time step in our simulations has virtually no effect on the spectral power of variability at periods from about 2--3 days and larger so that such a choice is appropriate for our goals. 

As illustrated by Eq.~\ref{eq:F} the variability of the flux $F(\lambda, t)$ is brought about by the time-dependence of the solid angles of active regions seen from the vantage point of observer, $\Omega_i$, and by the time-dependence of facular and spot fractions $\alpha_{i,F}(t)$ and $\alpha_{i,S}(t)$. The former is attributed to the evolution of magnetic features as well as to the rotation of the star and consequent change of the foreshortening factor. The latter is given by the difference between facular and spot lifetimes. For example, in the (rather unrealistic) case of the same lifetime of spot and facular components of an active region, their relative coverage  would not depend on time and consequently the time dependence of the contribution of active regions to stellar brightness will be solely determined by the variable solid angle $\Omega_i(t)$. 

All in all, the power spectra of stellar brightness variations depend on properties of stellar active regions and the viewing geometry. To better illustrate the important effects and individual roles of each of the involved parameters we start with considering a greatly simplified case of variability in Sect.~\ref{sect:spots} and add  more realism into our simulations in the subsequent sections (while still keeping the model  relatively simple).

\section{Stars with spots}\label{sect:spots}
In this section we examine stellar brightness variations due to spots, i.e. we put $\alpha_S(t)$ to 1 and $\alpha_F(t)$ to 0 in Eq.~\ref{eq:F}. 
We track a star during a time interval of 1600 days which roughly corresponds to the duration of 17 Kepler quarters, i.e. approximately the total duration of the Kepler mission. During this interval we let 300 spots emerge, each at a random point of time and in a random place within the activity belts on the stellar surface (see Sect.~\ref{sect:model}). Since we are interested in the impact of spot evolution on the profile of the power spectrum of stellar brightness variations and, in particular,  on the position of the inflection point, 
 for illustrative purposes we assume  that all spots have the 
 same growth and lifetimes, independently of their size. 
We also start with a simple case of spots emerging in three relative sizes scaling as 1:2:3 and consider 100 spots of each of the sizes. A more realistic treatment will be employed in the subsequent sections. 
We note that the absolute size of spots does not play a role in the calculations presented in this section since it affects only the amplitude of the brightness variations and has no effect on the profile of their power spectrum. 

\subsection{High-frequency tail of the power spectrum and inflection point}
\begin{figure}
\resizebox{\hsize}{!}{\includegraphics{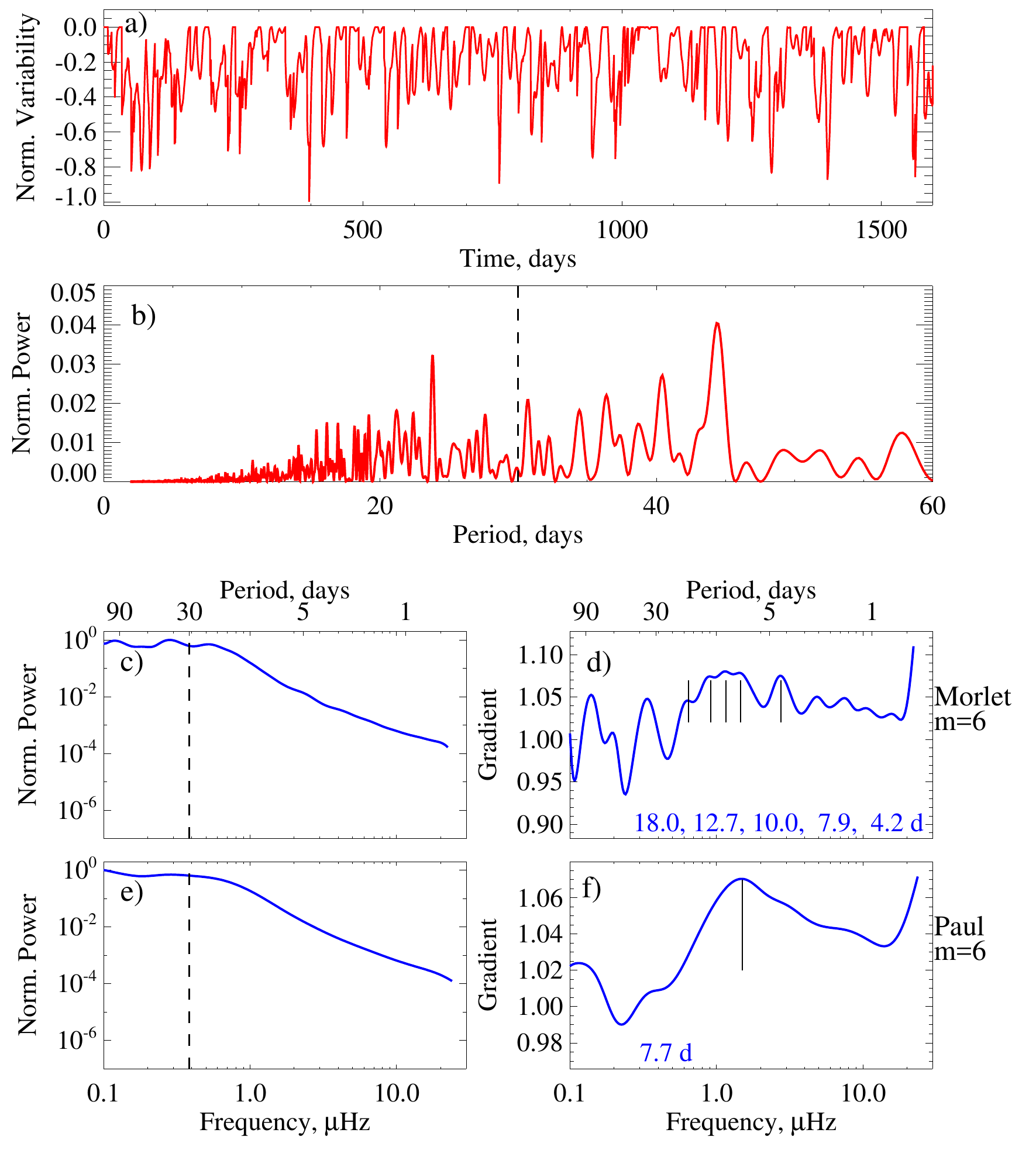}}
\caption{Model light curve of a star with a 30-day rotation period covered by spots and observed from its equatorial plane. The spots decay according to a {\it linear} law with $T_{\rm spot}=25$ d. Two upper panels show a normalized light curve (panel a) and corresponding Lomb-Scargle periodogram (panel b). Panels c--f show global wavelet power spectra (left panels) and corresponding gradient of the power spectra (right panels) calculated with the 6th order Morlet wavelet (panels c and d) and with the 6th order  Paul wavelet (panels e and f). The values of the gradients of these power spectra are a scaled and offset by unity (see Eq.~\ref{eq:RK} and discussion in the text for the exact quantity plotted). Numbers in panels d and f correspond to the positions of the inflection points (i.e. local maxima of the gradient). Vertical dashed lines in panels b, c, and e indicate the rotation period of the modeled star. Vertical solid lines in panels d and f indicate positions of the inflection points. We note that since spots reduce stellar brightness, the normalized variability (i.e. normalized $F(\lambda,t)-F_{Q} (\lambda)$ values, see Eq.~\ref{eq:F}) is plotted between -1 and 0.}
\label{fig:Ex1}
\end{figure}

\begin{figure}
\resizebox{\hsize}{!}{\includegraphics{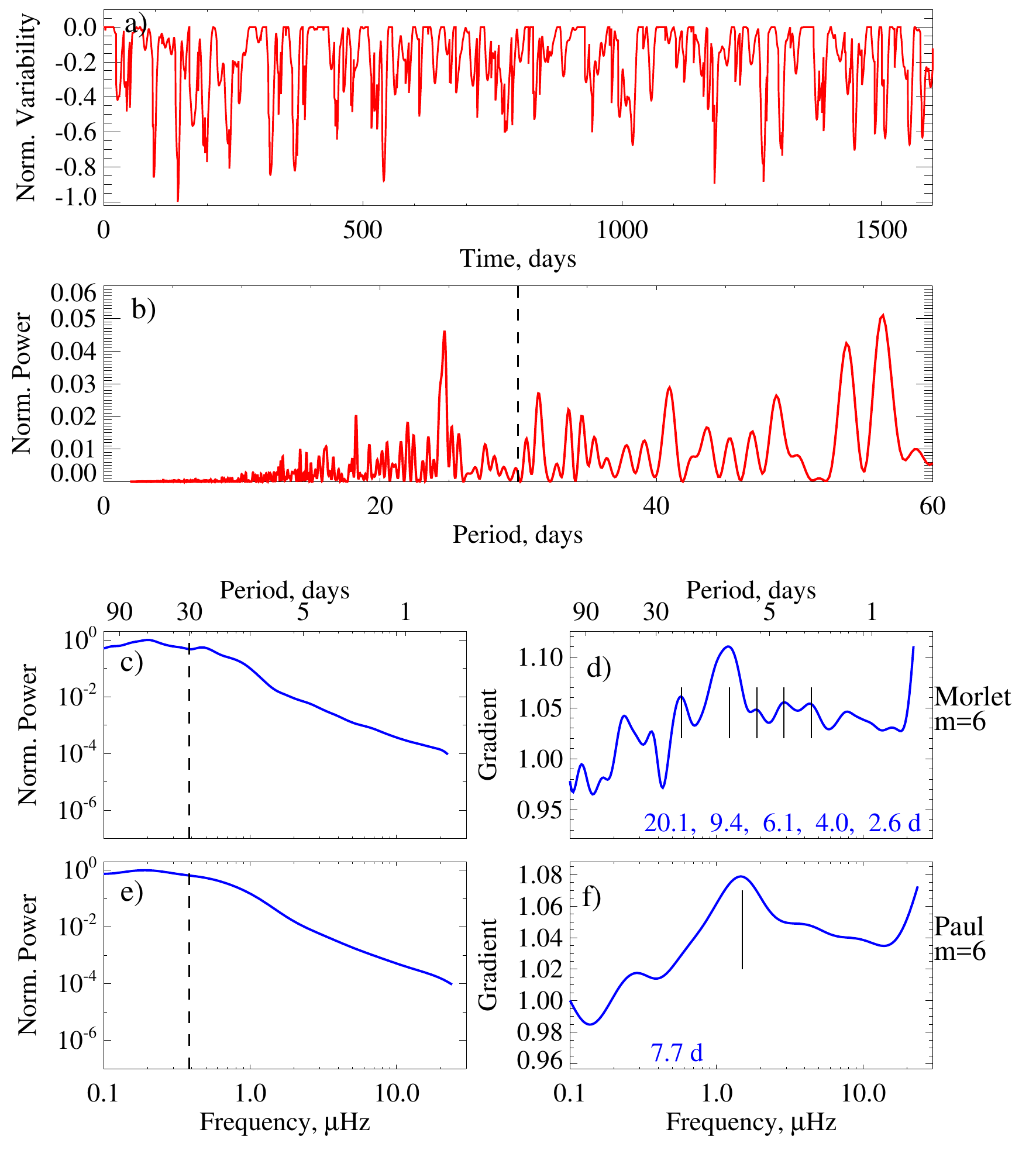}}
\caption{The same as Fig~\ref{fig:Ex1} but for another realization of spot emergences.}
\label{fig:Ex2}
\end{figure}

Figures~\ref{fig:Ex1}~and~\ref{fig:Ex2} show two realizations of light curves calculated for a model star rotating with a 30-day period. We assumed that spots instantaneously emerge on the stellar surface (i.e. that the growth time is zero) and then their areas linearly decrease with time. In other words, the spot area $A(t)$ after the emergence can be written as
\begin{equation}\label{eq:lin}
 A(t)=A_0 \left ( 1 - \frac{t-t_0}{T_{\rm spot}} \right ), \,\,\, t_0 \le t \le t_0+T_{\rm spot},
\end{equation}
where $A_0$ is the maximum area and $t_0$ is the time of emergence. We put $T_{\rm spot}=25$ d to produce light curves for Figs.~\ref{fig:Ex1}~and~\ref{fig:Ex2}. Since times and positions of individual emergences are kept random, the two light curves shown in these figures are distinctly different from each other. 

One can clearly see the individual dips caused by the transits of spots as a star rotates (Figs~\ref{fig:Ex1}a~and~\ref{fig:Ex2}a). Nevertheless, the Lomb-Scargle periodograms of both light curves do not have a clear 30-day peak (Figs~\ref{fig:Ex1}b~and~\ref{fig:Ex2}b). Instead, the peaks appear to be rather random and their locations depend on the specific realization of spot emergences. The same situation is seen when global wavelet power spectra with 6th order Morlet and Paul wavelets (see Figs~\ref{fig:Ex1}c~and~\ref{fig:Ex2}c and Figs~\ref{fig:Ex1}e~and~\ref{fig:Ex2}e, respectively) are computed: all four power spectra do not have any noticeable signature of the rotation peak. In comparison to the Morlet wavelet, Paul wavelet implies a poorer frequency localization but stronger averaging in the frequency domain when power spectra are computed. Consequently, wavelet power spectra calculated with the Paul wavelet have less details but they are more resistant to statistical noise \citep{wavelet}.

\begin{figure}
\resizebox{\hsize}{!}{\includegraphics{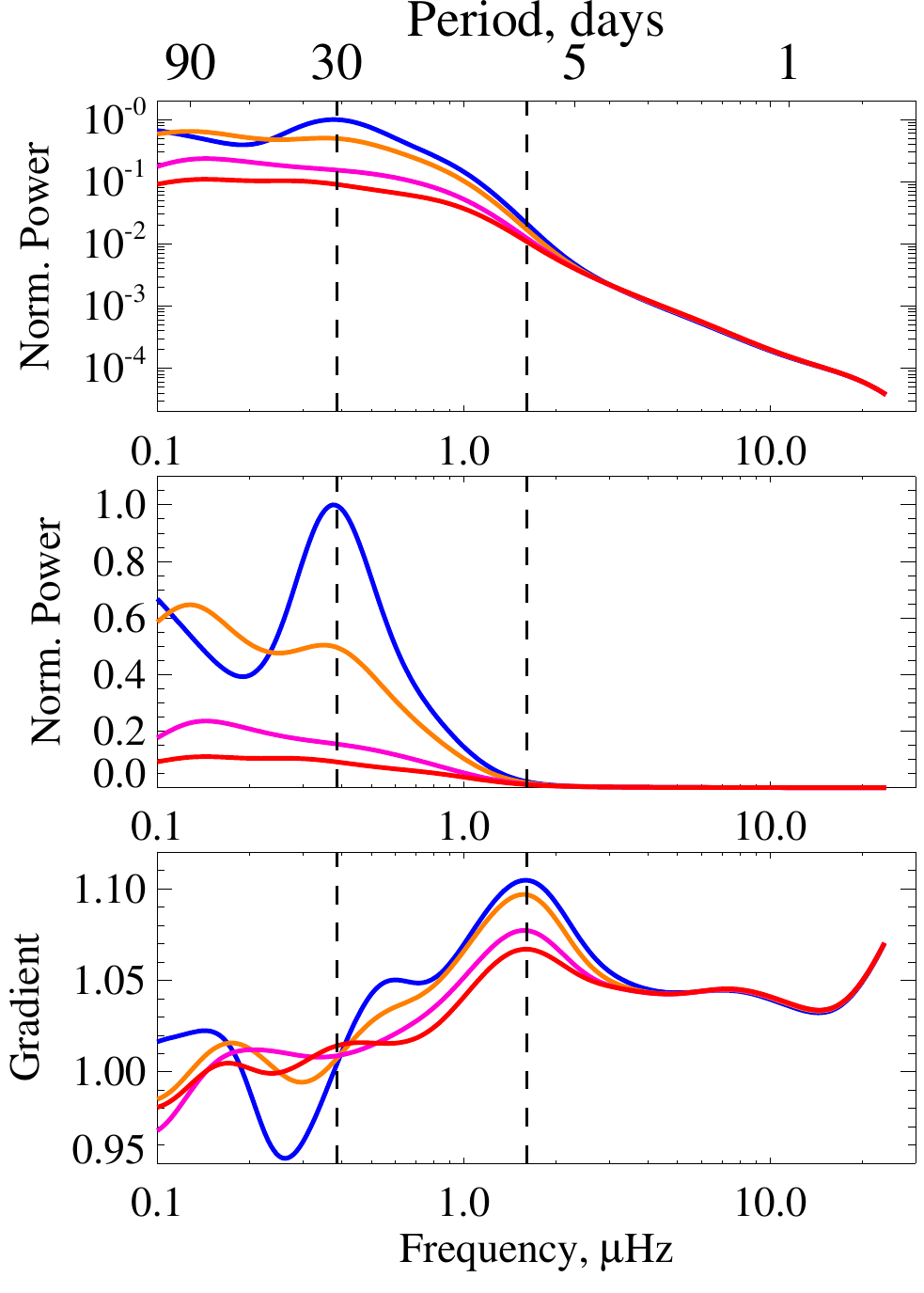}}
\caption{Power spectra of model light curves plotted on logarithmic (top panel) and linear (middle panel) scales on the vertical axis. The gradients of the power spectra in the top panel are plotted in the bottom panel. The modeled spots decay according to a {\it linear} law with lifetimes, $T_{\rm spot}$, equal to: 90 d (blue), 50 d (orange), 20 d (magenta), and 12 d (red). All four light curves are calculated for the same realization of spot emergences. Vertical dashed lines at 30 d and 7.2 d correspond to the rotation period of the simulated star and the approximate position of the inflection point in all four power spectra, respectively. Power spectra are calculated with the 6th order Paul wavelet.}
\label{fig:spot_dec}
\end{figure}

In Figs~\ref{fig:Ex1}d, f~and~\ref{fig:Ex2}d, f we plot the ratios $R_k$ between the power spectral density $P(\nu)$ at two adjacent frequency grid points: $R_k \equiv P(\nu_{k+1})/P (\nu_k)$. It is easy to show that these ratios can be written as 
\begin{equation}\label{eq:RK}
R_k = 1+\frac{d \ln P(\nu_k)}{d \ln \nu} \cdot \frac{{(\Delta \nu)}_k}{\nu_k},
\end{equation}
where $\Delta \nu$ is spacing of the frequency grid. We calculate power spectra on a  grid that is equidistant on a logarithmic scale, i.e. $\Delta \nu / \nu$ is constant. Therefore, $R_k$ values represent gradient of the power spectrum plotted on a log-log scale (as in Figs~\ref{fig:Ex1}c, e~and~\ref{fig:Ex2}c, e), scaled with some factor (which depends on the chosen frequency grid) and offset by unity. 

For simplicity from now on we will refer to the $R_k$ values as the gradient of the power spectrum. One can see that while the gradient of the Morlet power spectra has sophisticated profiles with many local maxima (corresponding to inflection points in the Morlet power spectrum), the gradient of the Paul power spectra looks much simpler.
Furthermore, while the power spectra of both light curves have no noticeable peak at the rotation period, both 6th order Paul power spectra have inflection  points giving rise to very clear peaks in the gradients of the power spectra. Importantly, the location of these points is the same for the two realizations plotted in   Figs.~\ref{fig:Ex1}~and~\ref{fig:Ex2}.



In Figs.~\ref{fig:Ex1}~and~\ref{fig:Ex2} we show two light curves corresponding to the same lifetime of spots, but to different realizations of spot emergences. In Fig.~\ref{fig:spot_dec} we look at things the other way around and consider power spectra of four light curves calculated with the same realization of spot emergences but with different lifetimes of spots. The power spectrum of the light curve with a spot lifetime $T_{\rm spot}=90$ days has a pronounced rotation peak. Its amplitude decreases rapidly with decreasing spot lifetime and disappears completely  when the lifetime of spots becomes smaller than the stellar rotation period: neither $T_{\rm spot}=20$ days nor $T_{\rm spot}=12$ days cases display any signature of the peak in the power spectrum around the rotation period. To better illustrate this point we also plot the power spectra on a linear vertical scale (Fig.~\ref{fig:spot_dec}b).

Figure~\ref{fig:spot_dec} illustrates  that stellar rotation periods cannot be determined from the maximum of the power spectrum when lifetimes of spots are small in comparison to the rotation period (at least for a star with no faculae). Interestingly, this is the case for the Sun since sunspots very rarely last longer than the solar rotation period \citep{BaumannSolanki2005}.

The bottom panel of Fig.~\ref{fig:spot_dec} points to an alternative method for determining rotation periods when spot lifetimes are shorter than the stellar rotation period. One can see that the high-frequency tail of the power spectra is much less sensitive to spot lifetime. In particular, the position of the inflection point is almost the same in all four cases. The results obtained so far strongly suggest that  the high-frequency tail of the power spectrum may provide a more robust way of determining  stellar rotation periods. There are different ways of parameterizing the tail, e.g. one can approximate it with the help of a multi-component powerlaw fit similar to that employed by \cite{Aigrainetal2004} and establish the connection  between parameters of such a fit and the rotation period. However, in the present study we limit ourselves to showing that the position of the inflection point is a sensitive proxy of the stellar rotation period, leaving other methods for future investigations. 

The profile of the power spectrum and, consequently, the calibration factor between the position of the inflection point and rotation period depend on the wavelet utilized for calculations. The wavelets with good frequency localization lead to power spectra with multiple, often many inflection points whose positions depend on the specific realization of emergences (compare Figs.~\ref{fig:Ex1}d~and~\ref{fig:Ex2}d). At the same time wavelets with very low frequency localization  lead to a strong scatter in the relationship between inflection point position and the rotation period. After considering several wavelets with different degrees of frequency localization we found that the 6th order Paul wavelet introduces the best smoothing of the power spectra for our purposes.  {An example of power spectra and corresponding gradients calculated utilizing wavelets with different frequency localization is given in  Fig.~\ref{fig:wavelet}.}

We stress that the inflection point itself does not have a clear physical meaning and it is just a convenient way of quantifying the profile of the high-frequency tail of the power spectrum. 


\begin{figure}
\resizebox{\hsize}{!}{\includegraphics{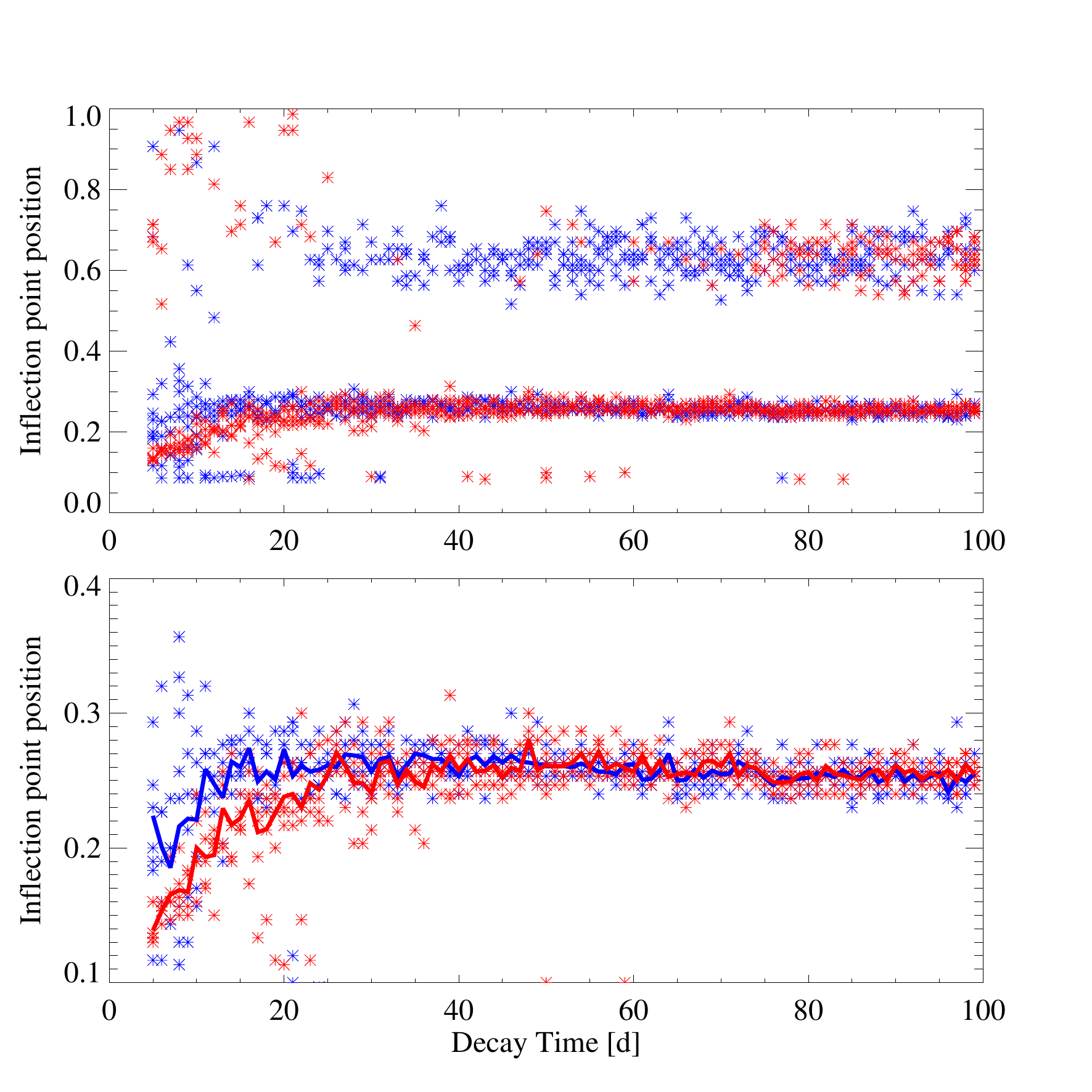}}
\caption{Dependence of the inflection point position (in fractions of the rotation period, $P_{\rm rot}=30$ d) on spot lifetime, $T_{\rm spot}$. Each value of the spot lifetime corresponds to five realizations of spot emergences with linear (red asterisks) and five realizations with exponential (blue asterisks) decay laws. Lower panel is a zoom in of the upper panel. Blue and red lines in the lower panel show positions of the high-frequency inflection points averaged over corresponding five realizations.}
\label{fig:exp_vs_lin}
\end{figure}

\begin{figure}
\resizebox{\hsize}{!}{\includegraphics{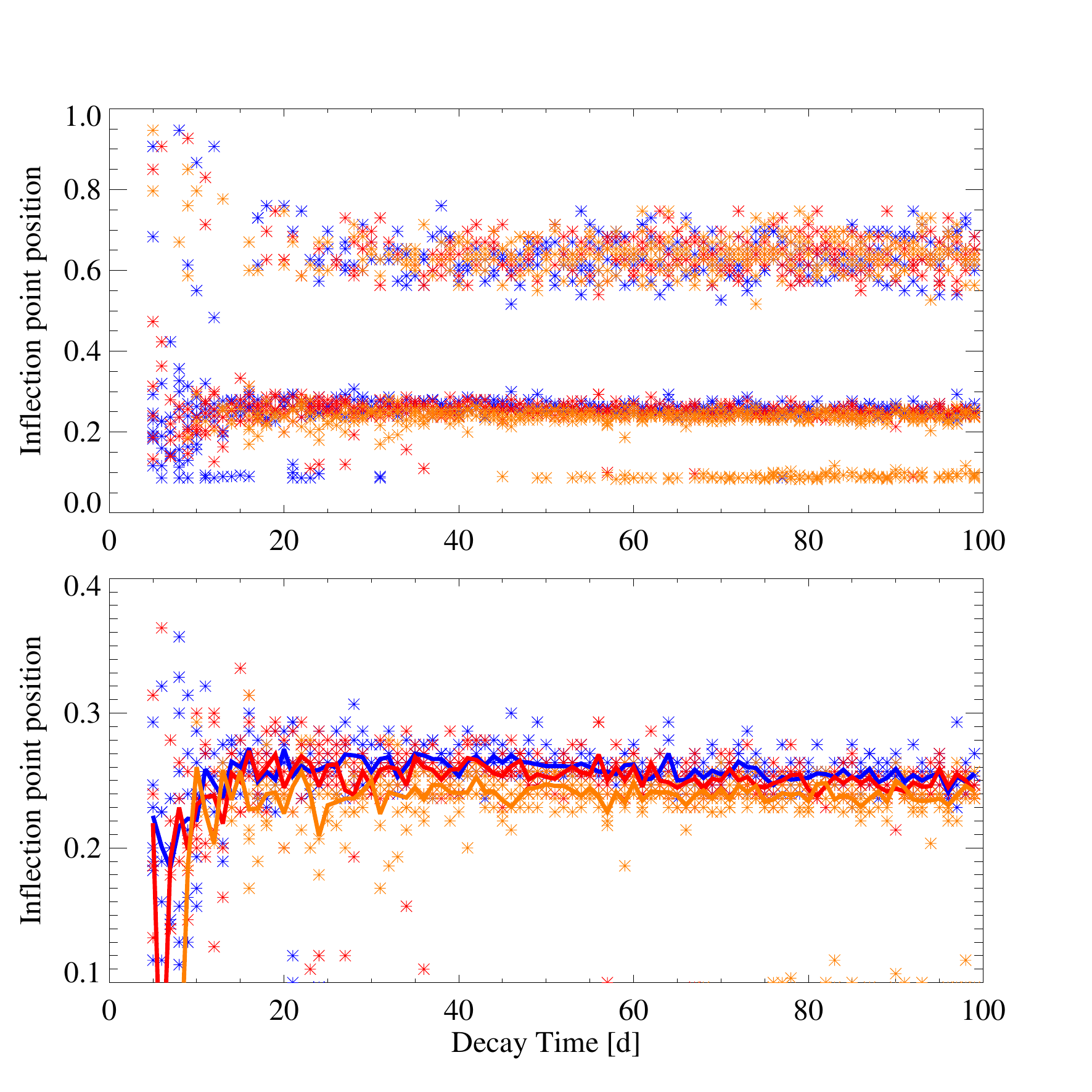}}
\caption{The same as Fig.~\ref{fig:exp_vs_lin}, but now comparing inflection points obtained for three values of spot emergence time: 0 d (blue), 1 d (red), and 2 d (orange). An exponential decay law is imposed. }
\label{fig:exp_em}
\end{figure}

\begin{figure*}
\resizebox{0.49 \hsize}{!}{\includegraphics{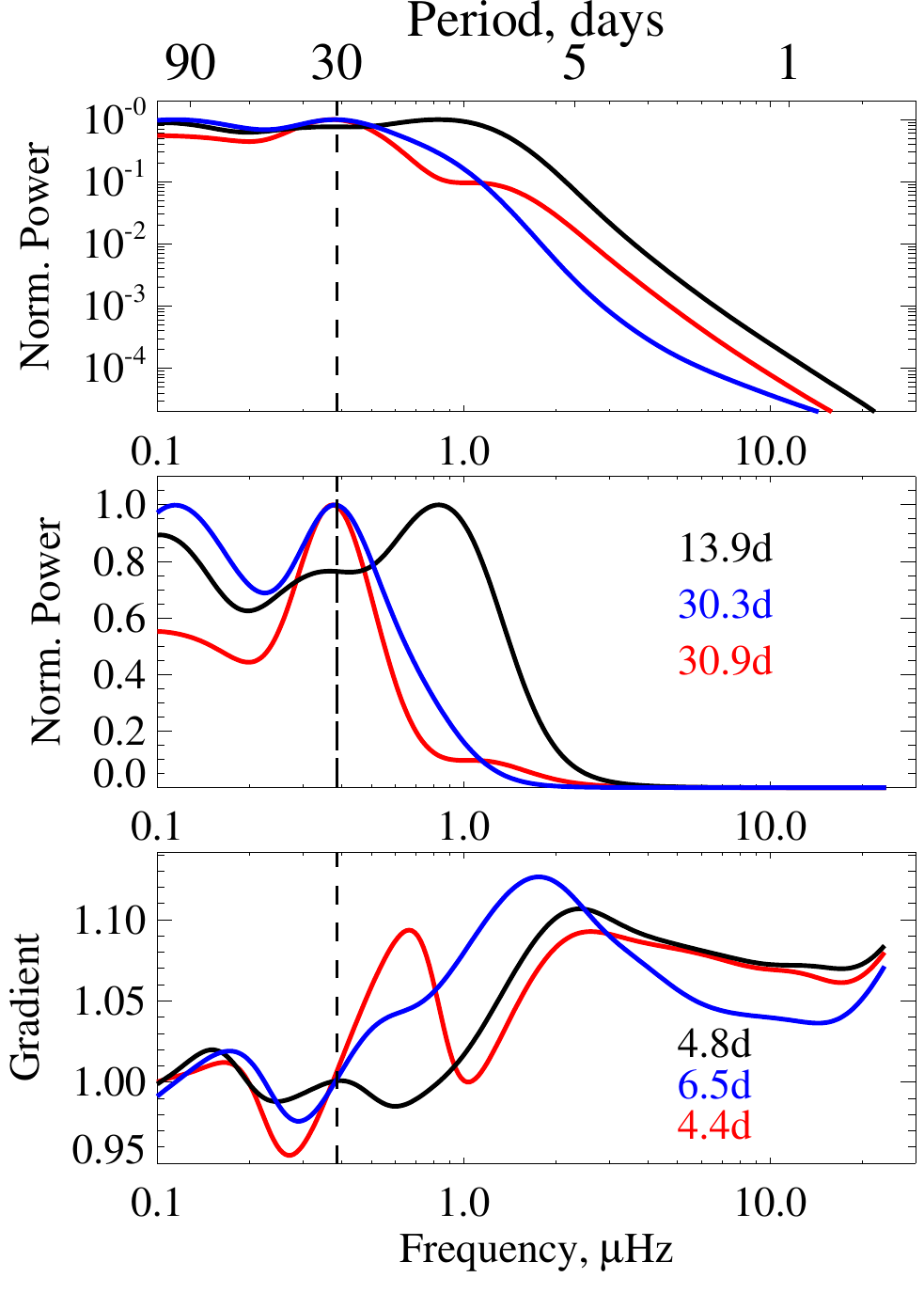}}
\resizebox{0.49 \hsize}{!}{\includegraphics{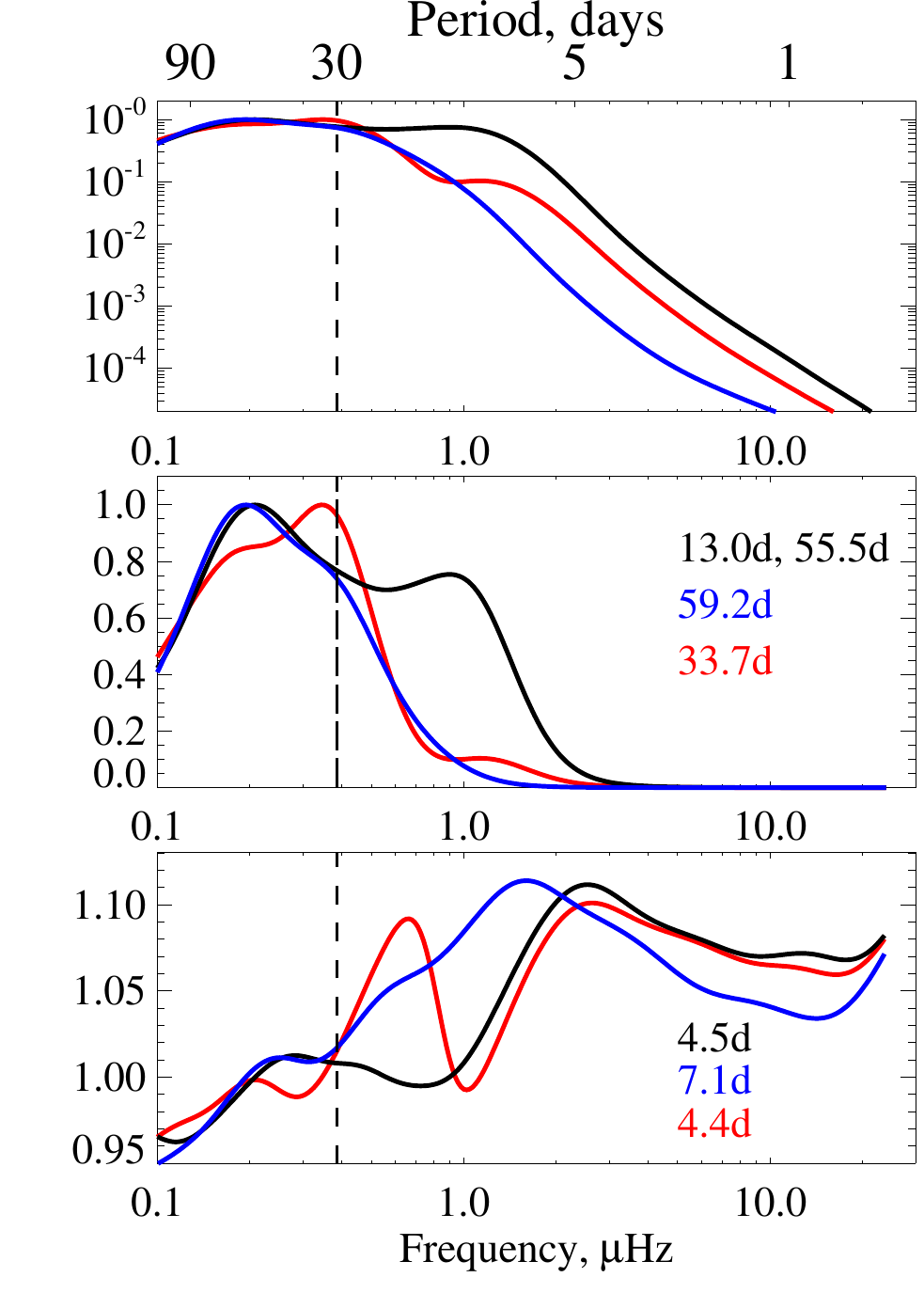}}
\caption{Power spectra of brightness variations of a star rotating with a 30-day period plotted on a logarithmic (upper panels) and linear (middle panels) scales. Also plotted are the gradients of the power spectra (lower panels). The calculations presented in left and right hand panels have been performed with the same set of model parameters (see text for details) but are associated 
with two different realizations of active regions emergences. Black, red, and blue curves correspond to total brightness variations (calculated with $S_{\rm fac}/S_{\rm spot}=5.5$), as well as their facular and spot components, respectively.   Numbers in the middle panels indicate peaks in the power spectra, while numbers in the lower panels point to positions of the inflection points.}
\label{fig:fac_ex}
\end{figure*}

\subsection{Effect of spot emergence and lifetime}\label{subsect:decay}
In Fig.~\ref{fig:exp_vs_lin} we show the dependence of the inflection point position on the lifetime of spots. In addition to the linear decay law, several other functional forms, such as parabolic and exponential decays \citep{decay1,decay2,decay3} have been proposed \citep[see also][for a review]{Sami_spots}. To illustrate the impact of the functional form of the decay law on the inflection point position we also consider an exponential law under which the spot area can be written as 
\begin{equation}
A(t) = A_0 \exp(-\frac{t-t_0}{T_{\rm spot}}), \,\,\, t \ge t_0. 
\label{eq:exp_dec}
\end{equation}

Figure~\ref{fig:exp_vs_lin} shows that the position of the inflection point remains stable and is not affected by  $T_{\rm spot}$ for values above  about 15 d (i.e. one half of the rotation period) for the linear decay law and 10 d (one third of the rotation period) for the exponential decay law. Note, however, the slightly different meanings of $T_{\rm spot}$  for the two decay laws. $T_{\rm spot}$  corresponds to spot lifetime for linear decay (i.e. the spot disappears when $T_{\rm spot}$  is reached), while $T_{\rm spot}$  in the exponential decay law implies an e-folding time. 

Since the exact profile of the power spectrum depends on the realization of spot emergences, the positions of the inflection point show some scatter around the mean values (solid lines in Fig.~\ref{fig:exp_vs_lin}) for a fixed lifetime. This scatter represents an intrinsic uncertainty of using the inflection point as a proxy for the rotation period. For some of the realizations ``rogue'' inflection points at lower periods appear (seen below 0.1 in the upper panel). Furthermore, inflection points are also found at large periods, even forming a high-period branch for spot lifetimes larger than about 30 days.  These points are linked to the rotation peak in the power spectra (which is only present if spot lifetime is large, see blue curves in Fig.~\ref{fig:spot_dec}).

Until now we considered spots that emerge instantaneously on the stellar surface and then decay. Such an assumption is reasonable for our purposes since the emergence and growth of spots takes significantly less time than the decay and rarely lasts longer than a few days \citep[see, e.g.][]{Driel_LR}. To estimate the effect of the non-zero growth time we compare in Fig.~\ref{fig:exp_em} the positions of the inflection points calculated for a spot emergence (and growth) time of 0 d (i.e. assuming instantaneous emergence as in Fig.~\ref{fig:exp_vs_lin}), 1 d, and 2 d. We assumed a linear growth of spot area during the emergence phase. One can see that including a non-zero emergence phase slightly shifts the inflection point to lower periods, but the effect is relatively small (on average 8\% for 2d emergence time).




\section{Stars with spots and faculae}\label{sect:fac}
Sunspots are generally parts of bipolar magnetic regions, which also harbor smaller magnetic elements. Ensembles of these magnetic elements form
 bright faculae \citep[see, e.g.][for reviews]{Sami_B, MPS_AA}. 
Faculae are present on late-type stars and play an important role in stellar photometric variability \citep[see, e.g. discussion in][]{Shapiroetal2016, witzkeetal2018, Timo2018}. For example, faculae dominate the variability over the course of magnetic activity cycles for old stars, like the Sun \citep{lockwoodetal2007, Radicketal2018}. They also significantly affect solar brightness variations on timescales of a few days \citep[][]{Shapiroetal2016, Sasha_NAT} and thus one can expect that the position of the inflection point in power spectra of stars similar to the Sun is affected by the facular contribution to stellar brightness variability. In this section we investigate the effect of faculae on the position of the inflection point in the power spectrum of stellar brightness variations.

\subsection{Treatment of faculae}\label{subsect:fac_tr}
Here we extend the model outlined in Sect.~\ref{sect:model} and describe the treatment of the facular contribution to stellar brightness variability.
Furthermore, we relax the assumption of the equal lifetime for all magnetic features adopted in Sect.~\ref{sect:spots} for illustrative purposes. Instead we consider a more comprehensive model of the decay of magnetic features.

For simplicity we limit ourselves to the case of the instantaneous emergence of active regions. This should not affect any of the conclusions drawn here since the duration of the emergence does not have a strong impact on the position of the inflection point (see Sect.~\ref{subsect:decay}). We assume that immediately after the emergence all magnetic regions have the same fractional coverage by spot and facular components. Consequently, we calculate the power spectrum of photometric variations and position of the inflection point as a function of the facular to spot area ratio {\it at the time of maximum area}, $S_{\rm fac}/S_{\rm spot}$. We note that for the case of the instantaneous emergence the time of maximum area coincides with the time of emergence.  Since facular and spot lifetimes are generally different the ratio {\it at the time of maximum area}, $S_{\rm fac}/S_{\rm spot}$, is not identical to the {\it instantaneous} (i.e. {\it snapshot}) ratio obtained at any random instance.

We adopt a solar log-normal distribution of spot sizes, taken from \cite{BaumannSolanki2005}, but only considering spots larger than 60 MSH (micro solar hemisphere). Consequently, the size of the spot component of each emerging magnetic region was randomly chosen following the \cite{BaumannSolanki2005} distribution. The log-normal distribution implies that while most of the spots have small sizes of about 100 MSH, every now and then huge spots with sizes of more than 3000 MSH appear. Then instead of considering a constant lifetime of all spots as we did in Sect.~\ref{sect:spots} we follow  \cite{decay2} and consider a constant decay rate
of spots. This results in a linear decay law with large spots living longer than small spots. The choice of the decay rate is not straightforward since it is rather poorly constrained even in the solar case. We will consider values between 10 MSH/day given by Gnevyshev-Waldmeier relation between sunspot sizes and lifetimes \citep{Waldmeier1955} and 41 MSH/day given by \cite{decay2}. In any case, as will be shown below, the position of the inflection point is basically independent of the decay rate. 

The lifetimes of spots are computed from spot areas and decay rates. To calculate the lifetime of faculae we assume a fixed ratio between lifetimes of facular and spot components of the active region (which implies a fix decay rate also for faculae). Since lifetimes of the facular component are usually significantly larger than those of spots \citep[see, e.g. reviews by][]{Sami_B, ARevolution} the active regions in our model emerge as a mixture of spot and facular regions, but then spend a significant part of their lifetimes as purely facular regions.

In our simplified parametric consideration of active region evolution  we do not directly account for the faculae brought about by the decay of spots. Faculae from sunspot decay imply a) underestimation of facular areas in our model; b) deviations of the facular decay law from linear. Point a) can be indirectly taken into account by the increase of the $S_{\rm fac}/S_{\rm spot}$ ratio (in other words facular area in this ratio represents not only the facular features emerging together with spots but also the product of the spot decay). We also do not expect that point b) can noticeably affect our calculations since the exact time evolution of magnetic features does not have a strong impact on the position of the inflection point. Recently \cite{Emre2018} performed more realistic calculations of magnetic flux emergence and surface transport in stars with various rotation periods. As a next step we plan to employ their results in our modeling.

\subsection{Superposition of spot and facular contributions to stellar brightness variability}\label{subsect:super}
In Fig.~\ref{fig:fac_ex} we depict power spectra of brightness variations brought about by faculae, by spots, and by their mixture (red, blue, and black lines, respectively). We have put spot decay rate to 25 MSH/day, i.e. roughly in between the estimates given by \cite{Waldmeier1955} and \cite{decay2}. The facular components of active regions were set to live twice as long as spot components. We have considered 1600-day light curves and let 2400 emergences randomly happen during this time. The absence of any clustering of emergences in time implies that the mean activity level of a star during the entire time of simulations stays the same, i.e. we do not consider activity cycles. 2400 emergences resulted in a mean fractional disk spot coverage being about 0.3\% (due to the adopted log-normal distribution of spot sizes the exact value slightly depend on the specific realization of emergences), which is a typical solar value around the activity maxima. 
\begin{figure*}
\resizebox{\hsize}{!}{\includegraphics{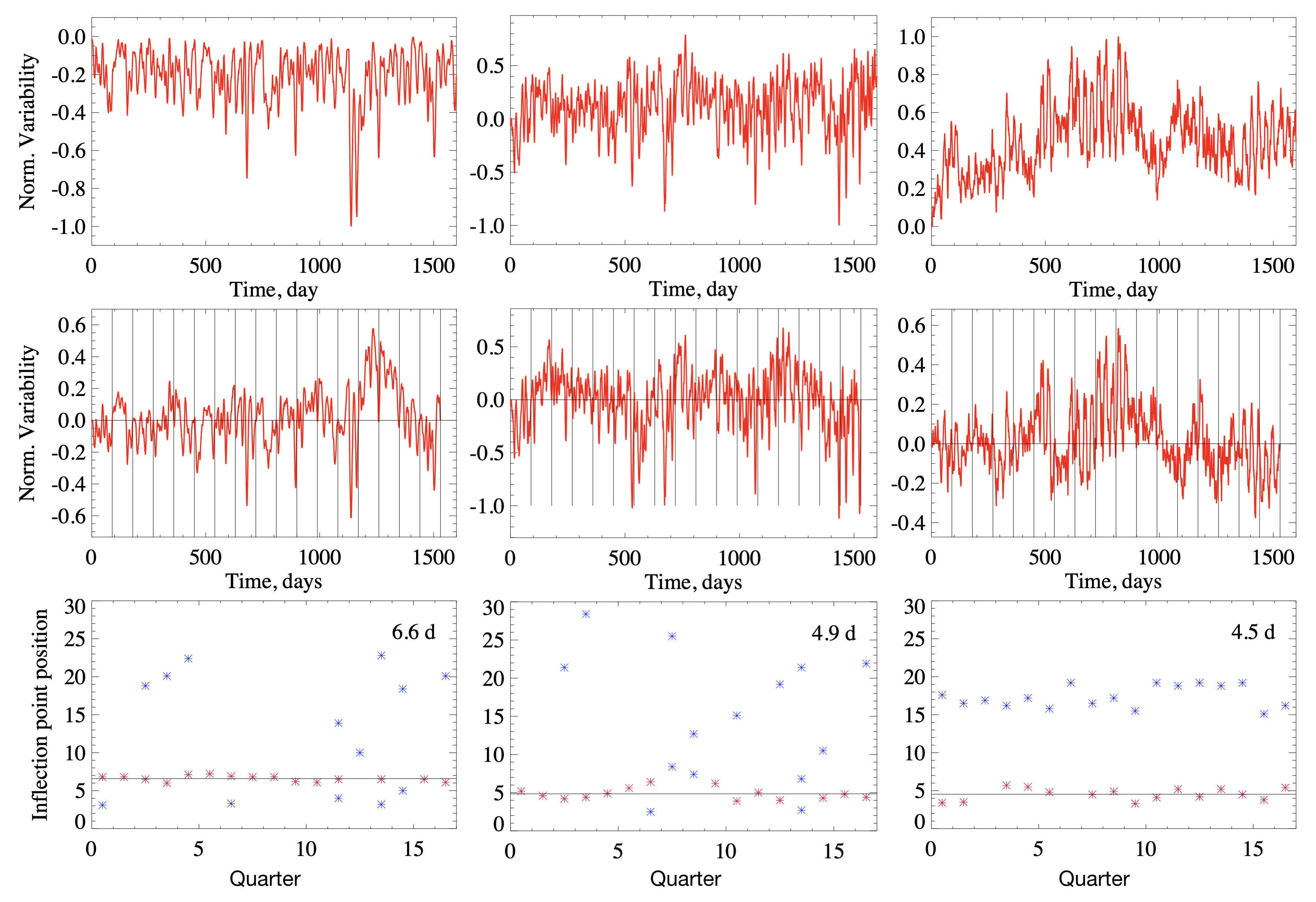}}
\caption{Three examples of simulated stellar variability: spot-dominated variability ($S_{\rm fac}/S_{\rm spot}=0.01$, left panels), intermediate case ($S_{\rm fac}/S_{\rm spot}=3$, middle panels), and faculae-dominated variability ($S_{\rm fac}/S_{\rm spot}=100$, right panels). Upper panels show original (i.e. without any detrending) light curves. Intermediate panels show light curves split in 17 90-day quarters and linearly detrended in each of the quarters. The separation between quarters is marked by the vertical black lines. The asterisks in the lower panels correspond to the positions of inflection points in each of the quarters. Numbers in the upper right corners of the 
lower panels are the outlier-resistant mean values of the inflection point positions. These values are also indicated in the lower panels by horizontal black lines. Red asterisks correspond to the inflection points utilized for calculating the outlier-resistant mean value, blue asterisks are trimmed as outliers.} 
\label{fig:example_fac}
\end{figure*}

Left and right panels of Fig.~\ref{fig:fac_ex} show power spectra of two light curves as well as of their facular and spot components. Both light curves have been calculated with the same set of model parameters specified above, but correspond to two different realizations of magnetic region emergences. In the realization plotted in the left panels both spot and facular components have a prominent peak at the stellar rotation period. However, since facular and spot components are in anti-phase at periods around the rotation period \cite[see discussion in][]{Sasha_NAT} the superposition of them leads to a disappearance of the rotation harmonic in the power spectrum of total brightness variations. Instead, a pronounced maximum in the power spectrum appears at 13.9 d, i.e. it is shifted by about 54\% from the rotation period. The bottom panels of Fig.~\ref{fig:fac_ex} shows that the location of the inflection point is different for the facular and spot components. This is not surprising since high-frequency tail of the power spectrum depends on the centre-to-limb variations of magnetic features contrasts and those are different for spots and faculae. In the given example the position of the inflection point of total brightness variations is shifted by 26\% relative to the position of the inflection point of the spot component alone. This number corresponds to the error in determining the rotation period which will be made if, in the absence of any information about the relative role of spot and facular components of the variability, one connects rotation period and position of the inflection point assuming purely spot-dominated variability. We note that in the case presented in the left panel of Fig.~\ref{fig:fac_ex} such an error is more than two times smaller than that made when assuming that rotation period corresponds to the maximum of the power spectrum (26\% vs. 54\%).

In the realization plotted in the right panels of Fig.~\ref{fig:fac_ex} the spot component does not have a maximum at the rotation period, while the facular component still shows a clear maximum (although slightly shifted to larger periods). 
In line with the discussion in Sect.~\ref{sect:spots} the position of the inflection point of the spot component is not affected by the disappearance of the peak corresponding to the rotation period (the shift of 0.6 d is within the scatter between different realizations of emergences, see Figs.~\ref{fig:exp_vs_lin}--\ref{fig:exp_em}). The superposition of the facular and spot components results in two peaks in the power spectra of total brightness variations, one at 13.0 d (i.e. shifted from the rotation period by 57\%) and another at 55.5 d (i.e. shifted by 85\%). Both numbers are  larger than the shift of the inflection point caused by the facular component which is equal to 37\%.

\section{Main factors affecting position of the inflection point}\label{sect:rel}
In this section we investigate the dependence of the inflection point position on the facular to spot area ratio {\it at the time of maximum area}, $S_{\rm fac}/S_{\rm spot}$, (Sect.~\ref{subsect:spot_fac}) and test this dependence against the solar case (Sect.~\ref{subsect:Sun}). We also establish the dependence of the inflection point position on stellar inclination (Sect.~\ref{subsect:incl}).

In Sect.~\ref{sect:spots}--\ref{sect:fac} we synthesized 1600-day light curves and then  employed them for calculating  power spectra and positions of the inflection points. While such a definition of the inflection point was appropriate for the illustrative purposes of Sect.~\ref{sect:spots}--\ref{sect:fac}, here we update the way the position of the inflection point is calculated to bring our calculations more into line with  available stellar photometric data (e.g. Kepler or TESS) 

As in Sect.~\ref{sect:spots}--\ref{sect:fac} we synthesize 1600-day light curves but instead of employing them directly for calculating  positions of the inflection points we first make ``Kepler-like'' light curves out of them. In other words, we split the light curves in 17 90-day quarters (ignoring the last 70 days) and linearly detrend each of the quarters. Then, instead of calculating the positions of the inflection points using the entire light curve, we calculate the positions of the inflection points in every quarter and consider the outlier-resistant mean, ignoring points outside of two  {standard} deviations from the mean value. 

This procedure is illustrated in Fig.~\ref{fig:example_fac} for spot- and faculae-dominated variability as well as for the intermediate case of the facular to spot area ratio {\it at the time of maximum area} (see Sect.~\ref{subsect:fac_tr}), $S_{\rm fac}/S_{\rm spot}=3$ (compare top and middle panels to see the difference between original and ``Kepler-like'' light curves). We have adopted a value of 25 MSH/day for the sunspot decay rate and set the facular lifetime to be three times that of spots. The positions of the inflection points in each of the quarters are plotted in the bottom panels of Fig.~\ref{fig:example_fac}. The inflection points cluster in branches and, in particular, one can clearly see the branch corresponding to the high-frequency inflection point (i.e. at about 5--7 d). In the case of faculae-dominated variability, there is also a stable branch of low-frequency inflection points (in between 15 and 20 d, see right bottom panel of Fig.~\ref{fig:example_fac}). This is due to the lifetime of faculae being sufficiently large for preserving a low-frequency inflection point (see discussion in Sect.~\ref{subsect:decay}). Since the high-frequency branch is more stable we constrain ourselves to its analysis and refrain from studying the low-frequency branch.  {The existence of the low-frequency branch might potentially lead to an ambiguity in the period determination. If, for example, the high-frequency branch is not visible due to the high noise level in the data then low-frequency branch might be erroneously taken for the high-frequency branch. This would lead to a roughly four times overestimation of the period. Such an ambiguity can be resolved by applying additional criteria.  For example, one would expect that rotational periods of fast rotators should be caught by the autocorrelation or Lomb-Scargle periodograms techniques. Consequently, if there is, for example, an ambiguity between rotation periods of 7 and 28 days and both Lomb-Scargle periodograms and autocorrelation analysis fail, then the 28-day value should be chosen for the rotation period.}


Figure~\ref{fig:example_fac} also shows that the positions of the inflection points slightly fluctuate from quarter to quarter and sporadically ``rogue'' inflection points appear. This is because the exact profile of the power spectrum depends on the specific realization of emergences of  magnetic regions.

\subsection{Position of the inflection point as a function of the facular to spot area ratio}\label{subsect:spot_fac}
\begin{figure}
\resizebox{\hsize}{!}{\includegraphics{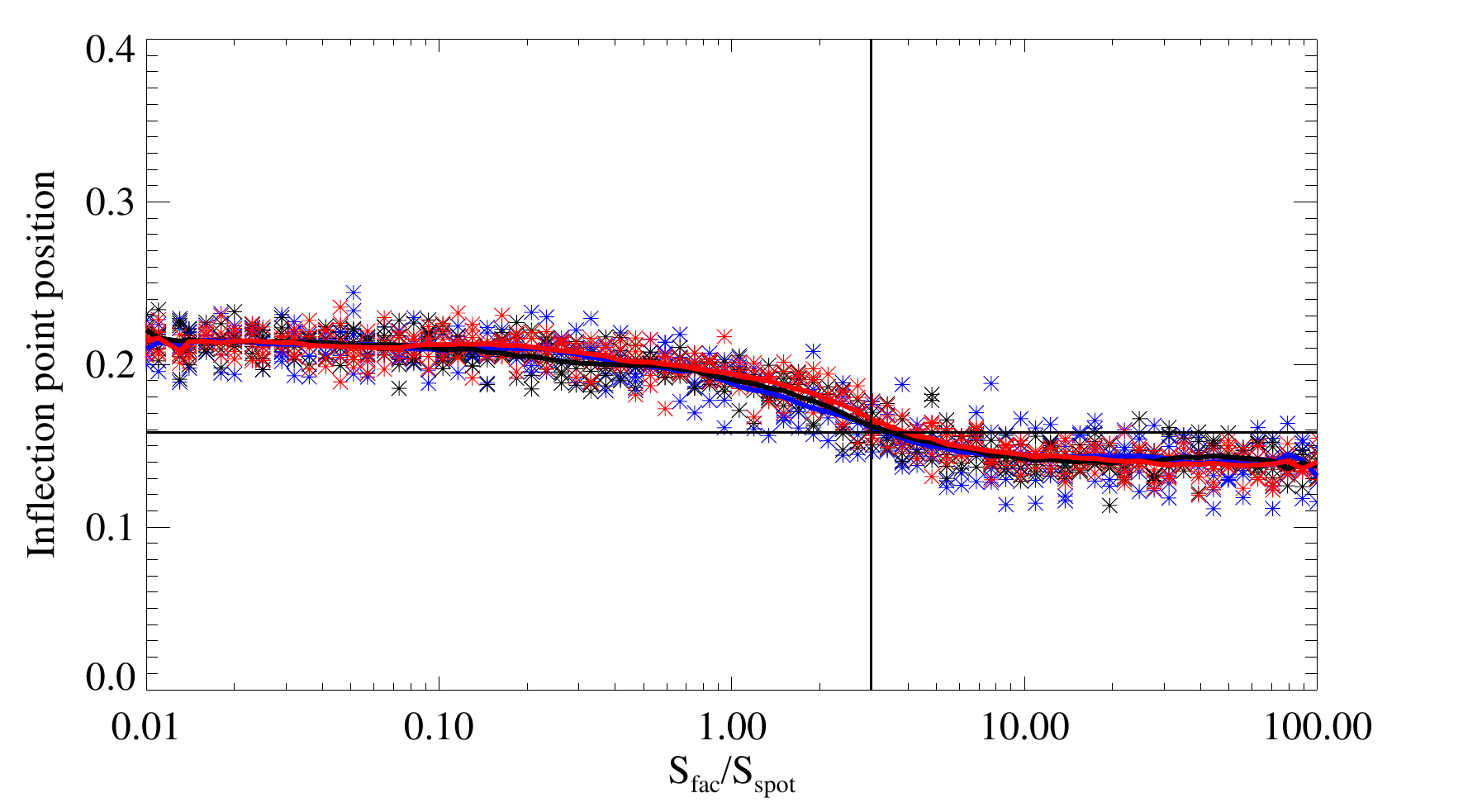}}
\caption{Dependence of the inflection point position (given as a fraction of the rotation period, $P_{\rm rot}=30$ d) on the facular to spot area ratio at the time of maximum area, $S_{\rm fac}/S_{\rm spot}$. Shown in red are calculations with lifetime of facular component of active regions, $T_{\rm fac}$, equal to the lifetime of spot component, $T_{\rm spot}$. Black and blue correspond to the $T_{\rm fac}/T_{\rm spot}=2$ and $T_{\rm fac}/T_{\rm spot}=3$ cases, respectively. For each pair of $S_{\rm fac}/S_{\rm spot}$ and $T_{\rm fac}/T_{\rm spot}$ values five realizations of emergences of active regions are shown. In other words, each of the $S_{\rm fac}/S_{\rm spot}$ values correspond to five red, five black, and five blue asterisks. Red, black, and blue lines mark the  positions of the inflection points averaged over the five corresponding realizations. The black horizontal line indicates the position of the solar inflection point from \cite{Eliana1}, while the black vertical line marks the  solar $S_{\rm fac}/S_{\rm spot}$ value established in Appendix~\ref{sect:fac_to_spot}.}
\label{fig:fac_spot}
\end{figure}

In Fig.~\ref{fig:fac_spot} we present the dependence of the inflection point position on the area ratio between the facular and spot components of active regions {\it at the time of maximum area}, $S_{\rm fac}/S_{\rm spot}$. We keep the mean fractional disk-area spot coverage constant and set it to about 0.3\% (see Sect.~\ref{subsect:super}). Hence, the $S_{\rm fac}/S_{\rm spot}$ value affects only the facular coverage. 

The decay rate of spots was chosen to be 10 MSH/day, i.e. according to the Gnevyshev-Waldmeier relation. In agreement with the calculations presented in Figs.~\ref{fig:fac_ex}--\ref{fig:example_fac} we have considered a fixed ratio between lifetimes of facular and spot components of active regions ($T_{\rm fac}$ and $T_{\rm spot}$, respectively). We note that in the solar case the faculae last significantly longer than spots \citep[see, e.g., review by][]{Sami_B}. For example, \cite{premingeretal2011, Thierry2018} found that facular features can affect solar UV irradiance (where it can be  disentangled from noise more easily than in the white light) for up to 3--4 solar rotations (see their Fig.~5). In this context, Fig.~\ref{fig:fac_spot} shows calculations for $T_{\rm fac}/T_{\rm spot}=1$, $T_{\rm fac}/T_{\rm spot}=2$, and $T_{\rm fac}/T_{\rm spot}=3$ cases.

Figure~\ref{fig:fac_spot} shows that for spot-dominated variability (i.e. for small $S_{\rm fac}/S_{\rm spot}$ values) the inflection point is located at 22\% of the rotation period (we only plot the high-frequency inflection points). We note that a small shift with respect to the 25\% value seen in Figs.~\ref{fig:exp_vs_lin}--\ref{fig:exp_em} is brought about by the different procedures for calculating the inflection point position, i.e. taking the outlier-resistant mean of 17 90-day intervals instead of computing a single inflection point. In the case of faculae-dominated variability (i.e. of large $S_{\rm fac}/S_{\rm spot}$ values) the inflection point is located at about 14\% of the rotation period. The level of the  statistical noise (i.e. variations in inflection point position caused by the random pattern of active regions emergences) is about 2--3\%. 

While the position of the inflection point strongly depends on the $S_{\rm fac}/S_{\rm spot}$ value, the difference between the three considered  $T_{\rm fac}/T_{\rm spot}$ values is barely visible (compare red, blue, and black curves in Fig.~\ref{fig:fac_spot}). This has two important implications. First, auspiciously, the ambiguities in facular lifetime do not have a strong effect on the calculations of the inflection point position. Second, the position of the inflection point depends rather on the facular to spot area ratio {\it at the time of maximum area} than on the {\it instantaneous} ratio (which is proportional to the product of area ratio at the time of maximum area and ratio of the facular and spot lifetimes). We note that this result is in line with the discussion in Sect.~\ref{sect:spots}, where we showed that the position of the inflection point only weakly depends on the lifetime of magnetic features. 

Since stellar $S_{\rm fac}/S_{\rm spot}$ values are a priori unknown, their effect on the relation between rotation period and inflection point position introduces additional uncertainty in the period determined with the help of the inflection point (see Sect.~\ref{subsect:activity} for a more detailed discussion). At the same time the dependence of the inflection point position on the $S_{\rm fac}/S_{\rm spot}$ value makes it possible to determine the ratio for stars with known rotation periods. We note that since the dependence presented in Fig.~\ref{fig:fac_spot}  is rather noisy, it is more suitable for studying general trends (e.g. the dependence of facular to spot ratio on stellar activity) than for deducing $S_{\rm fac}/S_{\rm spot}$ values for individual stars. We plan to 
determine $S_{\rm fac}/S_{\rm spot}$ values for \cite{McQuillan2013} sample of 34,030 stars with known rotation periods in a forthcoming publication.  {In this paper we limit ourselves to giving an example of application of the GPS method to stars significantly more variable than the Sun  with presumably spot-dominated variability  (see  Appendix~\ref{Examples}).}

The calculations presented so far in this section have been performed for a fixed values of the rotation period, mean fractional disk-area spot coverage, and spot decay rates. In Appendix~\ref{extra} we illustrate that the calibration factor between the inflection point position and rotation period is only marginally influenced by the rotation period (Fig.~\ref{fig:app_period}), spot coverage (Fig.~\ref{fig:app_coverage}), and spot decay rate (Fig.~\ref{fig:app_decay}). Furthermore, we show that the position of the inflection point only barely depends on the latitude of the emerging active regions (Fig.~\ref{fig:app_lat}).

\subsection{Inflection point in the power spectrum of solar brightness variations}\label{subsect:Sun}
Let us now locate the Sun in Fig.~\ref{fig:fac_spot}. This requires knowledge of the inflection point position in the power spectrum of solar brightness variations as well as of the solar $S_{\rm fac}/S_{\rm spot}$ value. \cite{Eliana1} demonstrated that the inflection point in the power spectrum of solar brightness variations is located at a period of about 4.17 days which is roughly 15.9\% of the solar synodic rotation period at the equator. There have been also a number of studies aimed at determining the {\it instantaneous} ratio between facular and spot solar disk-area coverages \citep[see, e.g.][]{chapman1997}. At the same time the solar value of the facular to spot ratio {\it at the time of maximum area}, $S_{\rm fac}/S_{\rm spot}$ is, on the whole, rather poorly studied and until now has remained unknown. 
In Appendix~\ref{sect:fac_to_spot} we present a new method for determining the solar $S_{\rm fac}/S_{\rm spot}$ value and show that mean solar value over the 2010--2014 period is about 3. Fig.~\ref{fig:fac_spot} demonstrates that this value, in combination with the position of the solar inflection point from \cite{Eliana1}, agrees well with our calculations. This is reassuring, since it indicates that our simple parametric model allows accurate calculations of the inflection point position.

We remind that due to the lack of constraints on the dependence of $S_{\rm fac}/S_{\rm spot}$ value on size of magnetic regions we assumed the same $S_{\rm fac}/S_{\rm spot}$ ratio for all emerging magnetic regions. Solar data indicate that the {\it instantaneous} ratio between disk-area coverages by faculae and spot decreases from minimum to maximum of solar activity \citep{chapman1997,foukal1998, solankiandunruh2012, Shapiro2014_stars}. One can speculate that such a behavior is partly attributed to a stronger cancellation of small magnetic flux concentrations (associated with faculae) at higher levels of solar activity when regions with opposite polarities lie closer to  each other (Cameron 2018, private communication). 
 {Based on this one can suggest that the ratio {\it at the time of maximum area} should not show as strong dependence on solar activity  as the {\it instantaneous} ratio. This is in line with the results of \cite{Eliana1}, who could not pinpoint any clear dependence of the solar inflection point (which depends on the ratio {\it at the time of maximum area}, see above) on the level of solar activity. A possible changes of this ratio within a stellar activity cycle would contribute to the scatter in position of the inflection points.}

\subsection{Effect of inclination}\label{subsect:incl}
The trajectories of  active regions across the stellar disk as a star rotates depend on the position of the observer relative to the stellar equator. Consequently, stellar brightness variability is a function of the inclination \citep{schatten1993,knaacketal2001,luis2012,Shapiroetal2016}, which is the angle between the stellar rotation axis and the direction to the observer. Therefore, one can expect that the position of the inflection point depends on the inclination.

\begin{figure}
\resizebox{\hsize}{!}{\includegraphics{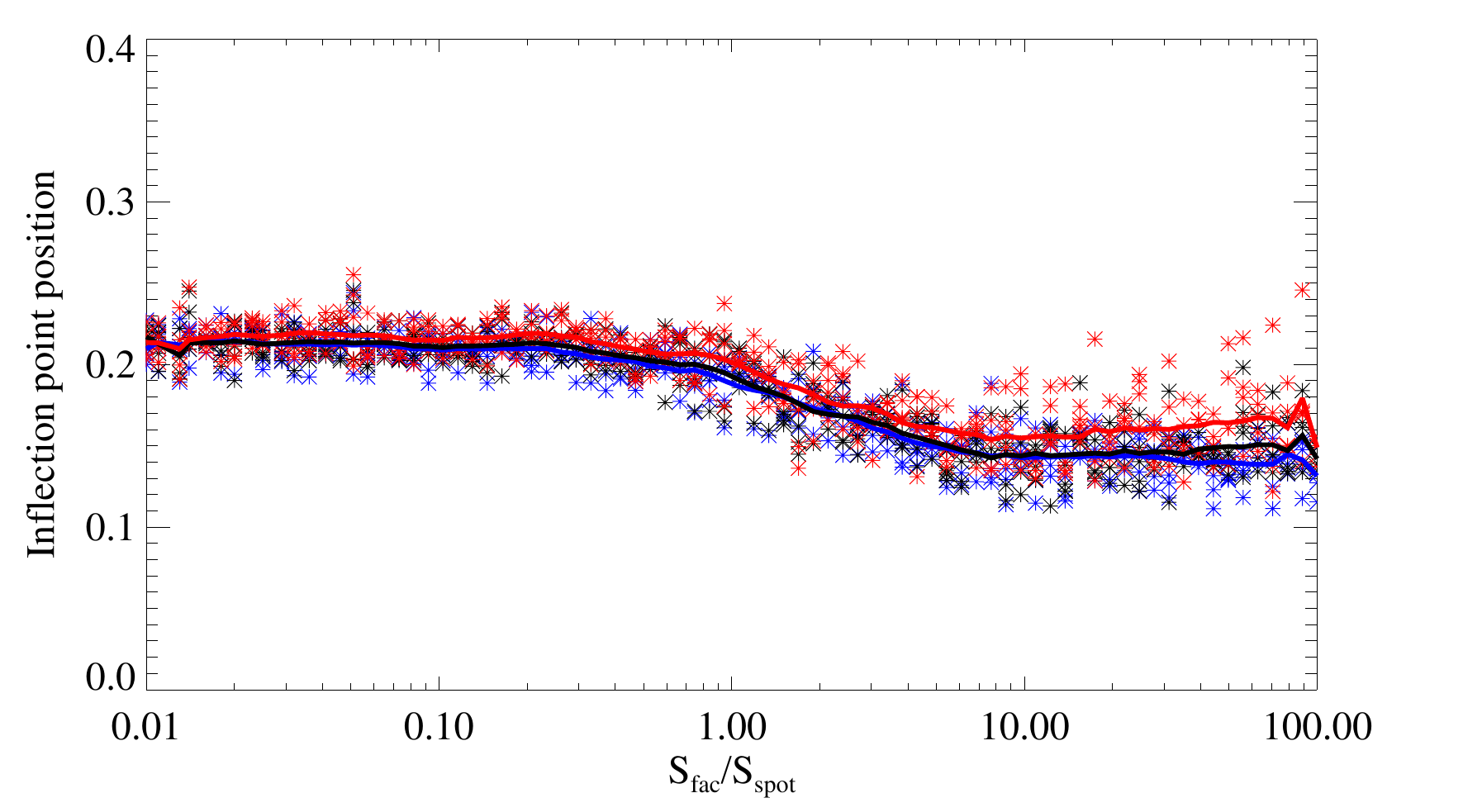}}
\caption{Sensitivity of inflection point to inclination of stellar rotation axis. Plotted is the dependence of the inflection point position (given as a fraction of the rotation period, $P_{\rm rot}=30$ d) on the ratio between facular and spot disk-area coverages at the time of maximum area, $S_{\rm fac}/S_{\rm spot}$. Each $S_{\rm fac}/S_{\rm spot}$ value corresponds to five realizations calculated with inclination $\varphi=90^{\circ}$ (equatorial view, blue), $\varphi=57^{\circ}$ (black), and $\varphi=45^{\circ}$ (red).
Red, black, and blue lines show positions of the inflection points averaged over the five corresponding realizations.}
\label{fig:fac_spot_incl}
\end{figure}

Figure~\ref{fig:fac_spot_incl} is the same as Fig.~\ref{fig:fac_spot}, except the different colored symbols now represent different stellar inclinations.
In contrast to Fig.~\ref{fig:fac_spot}, all calculations shown in Fig.~\ref{fig:fac_spot_incl} are performed with $T_{\rm fac}/T_{\rm spot}=3 $, but with three different values of the inclination: $90^{\circ}$ (blue), $57^{\circ}$ (black), and $45^{\circ}$ (red). An inclination of  $90^{\circ}$ corresponds to observations from the equatorial plane (so that the  blue asterisks are identical in Figs.~\ref{fig:fac_spot} and \ref{fig:fac_spot_incl}). An inclination of $57^{\circ}$ is the mean value of the inclination 
for a random distribution of rotation axes orientations.  One can see that all three dependences are very close to each other. Noticeable deviations in the inflection point position happen only for faculae-dominated stars with inclination value of $45^{\circ}$ (red asterisks in the right part of Fig.~\ref{fig:fac_spot_incl}).



\section{Position of the inflection point as a function of stellar magnetic activity}\label{subsect:activity}
The main goal of this section is to connect the position of the inflection point with proxies of stellar magnetic variability, namely with the S-index and photometric variability. When the facular to spot area ratio {\it at the time of maximum area}, $S_{\rm fac}/S_{\rm spot}$, is fixed, the position of the inflection point does not show any dependence on the total coverage of stellar surface by active regions (see Fig.~\ref{fig:app_coverage}). At the same time the level of magnetic activity affects the relative areas of facular and spot parts of active regions \citep{Shapiro2014_stars} and, consequently, the value of $S_{\rm fac}/S_{\rm spot}$. This leads to the dependence of the inflection point position on the magnetic activity which we study in this section.


In this context, we have simulated light curves with a different number of active regions emerging on each underlying star over the 1600-day period of simulations. We start with 80 emergences for the ``quietest'' light curves and end with 81000 emergences for the most ``active'' light curves. The sizes of spot components of active regions have been randomly chosen according to the log-normal distribution from \cite{BaumannSolanki2005} (see Sect.~\ref{subsect:fac_tr}). For each of the simulations we have calculated the mean value of the spot disk-area coverage and employed Eq. (1) from \cite{Shapiro2014_stars} to get the corresponding value of the S-index. Next we employed Eq. (2) from \cite{Shapiro2014_stars} to obtain the value of the facular disk-area coverage from the S-index. We have corrected this value by subtracting facular coverage corresponding to the absence of spots \citep[0.5\% according to Eqs. (1--2) from][]{Shapiro2014_stars}. Then we have calculated $S_{\rm fac}/S_{\rm spot}$ value, i.e. the ratio at the peak area of the active region,  which would result in such an instantaneous facular disk-area coverage. We have considered the $T_{\rm fac}/T_{\rm spot}=3$ case and set the decay rate of spots to 10 MSH/day (see Sect.~\ref{subsect:spot_fac}). 

\begin{figure}
\resizebox{\hsize}{!}{\includegraphics{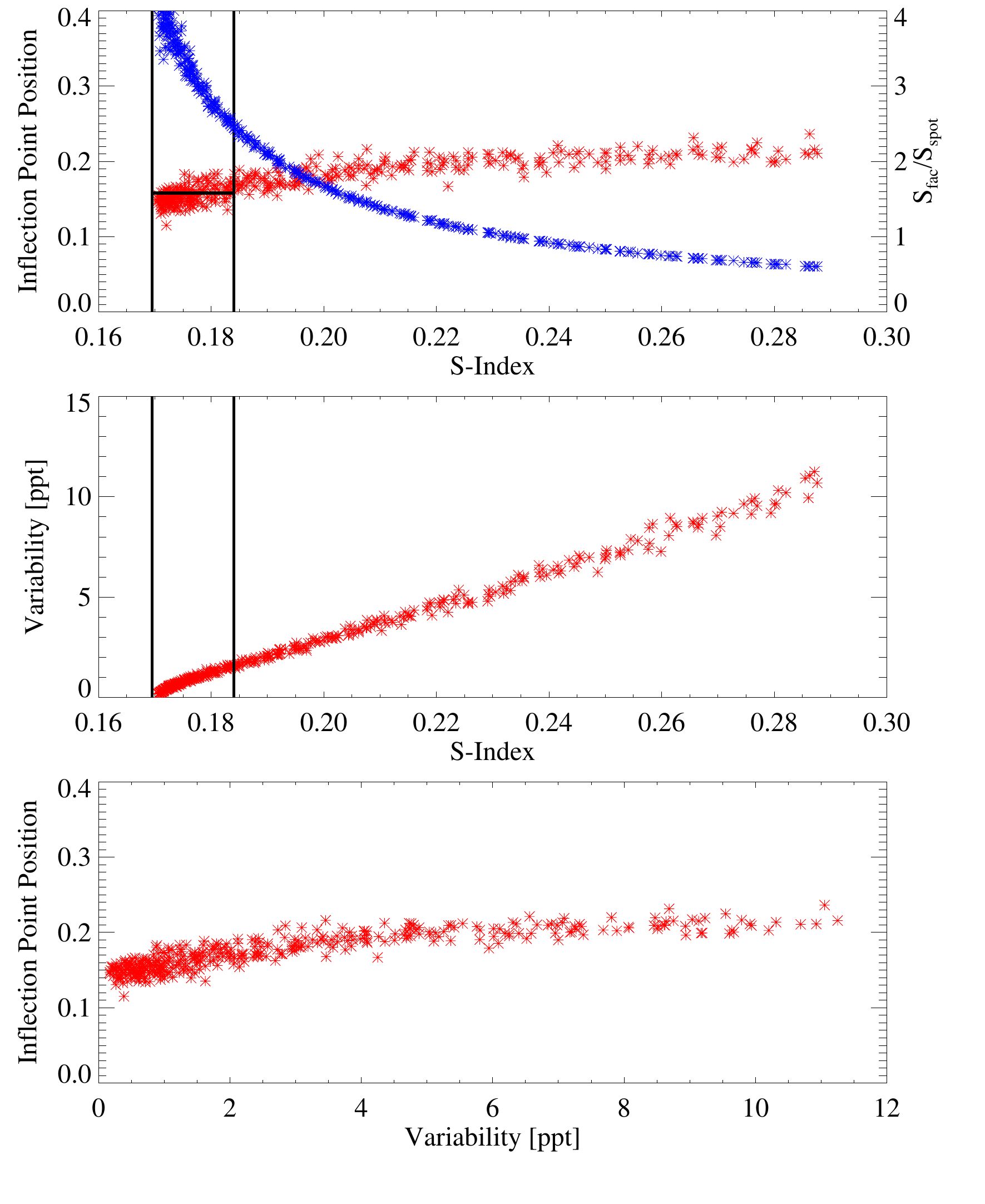}}
\caption{Dependence of the inflection point position (given as a fraction of the rotation period, $P_{\rm rot}=30$ d) on the S-index and stellar photometric variability (shown in red in top and bottom panels, respectively) as well as the dependence of the photometric variability on the S-index (middle panel). Blue asterisks in upper panel indicate the dependence of the facular to spot ratio at the time of maximum area, $S_{\rm fac}/S_{\rm spot}$, on the S-index. Black vertical lines in upper and middle panels point to the range of solar S-index values, while the horizontal black line in the top panel corresponds to the position of solar inflection point from \cite{Eliana1}. }
\label{fig:infl_act}
\end{figure}

Resulting dependences of the inflection point position and $S_{\rm fac}/S_{\rm spot}$ value on the S-index are given in the upper panel of Fig.~\ref{fig:infl_act}. One can see that the $S_{\rm fac}/S_{\rm spot}$ value decreases with the S-index. This is because Eqs. (1--2) in \cite{Shapiro2014_stars}  {are} based on the extrapolation from the solar case, where spot disk-area coverage depends on the S-index quadratically, while the dependence of facular disk-area coverage is linear. A decrease of the $S_{\rm fac}/S_{\rm spot}$ value with the S-index causes a rather weak shift of the inflection point to higher frequencies. For example, one can see that the position of the inflection point slightly shifts from solar minimum to solar maximum. At the same time the shift is smaller than the fluctuations of the inflection point caused by the statistical noise so that it does not contradict the results of \cite{Eliana1} (see Sect.~\ref{subsect:Sun}). Interestingly, the position of the inflection point remains  similar to that of the Sun even for significantly more active stars.

For each of the simulated light curves we calculate variability following the definition of variability range by \cite{Basri2011}. Namely, we split the light curves into 30-day segments. We sorted the segments by brightness and calculated the range between the 5th and 95th percentile of the brightness. Then we calculated the mean range among all 30-day segments. The resulting variability values are plotted in the middle panel of Fig.~\ref{fig:infl_act} as a function of the S-index. One can see that although the spot disk-area coverage increases quadratically with the S-index, the increase of the photometric variability is almost linear. This is because the variability range depends not on the absolute value of stellar disk-area coverages by active regions but rather on its fluctuations with time. The rise in the amount of active regions leads to a more uniform surface distribution which, in turn, decreases the variability range. 

Middle panel of Fig.~\ref{fig:infl_act} shows that solar variability range changes from almost zero during the solar minimum to roughly 1.5 ppt (parts per thousand). This agrees with a more accurate calculation in \cite{Shapiroetal2016} (see their Fig.~10a.). In the lower panel of Fig.~\ref{fig:infl_act} we plot the dependence of the inflection point position on the variability range. In most of the cases the position of the inflection point remains in between roughly 14\% and 21\% of the rotation period, even for stars significantly more variable than the Sun. 

\begin{figure}
\resizebox{\hsize}{!}{\includegraphics{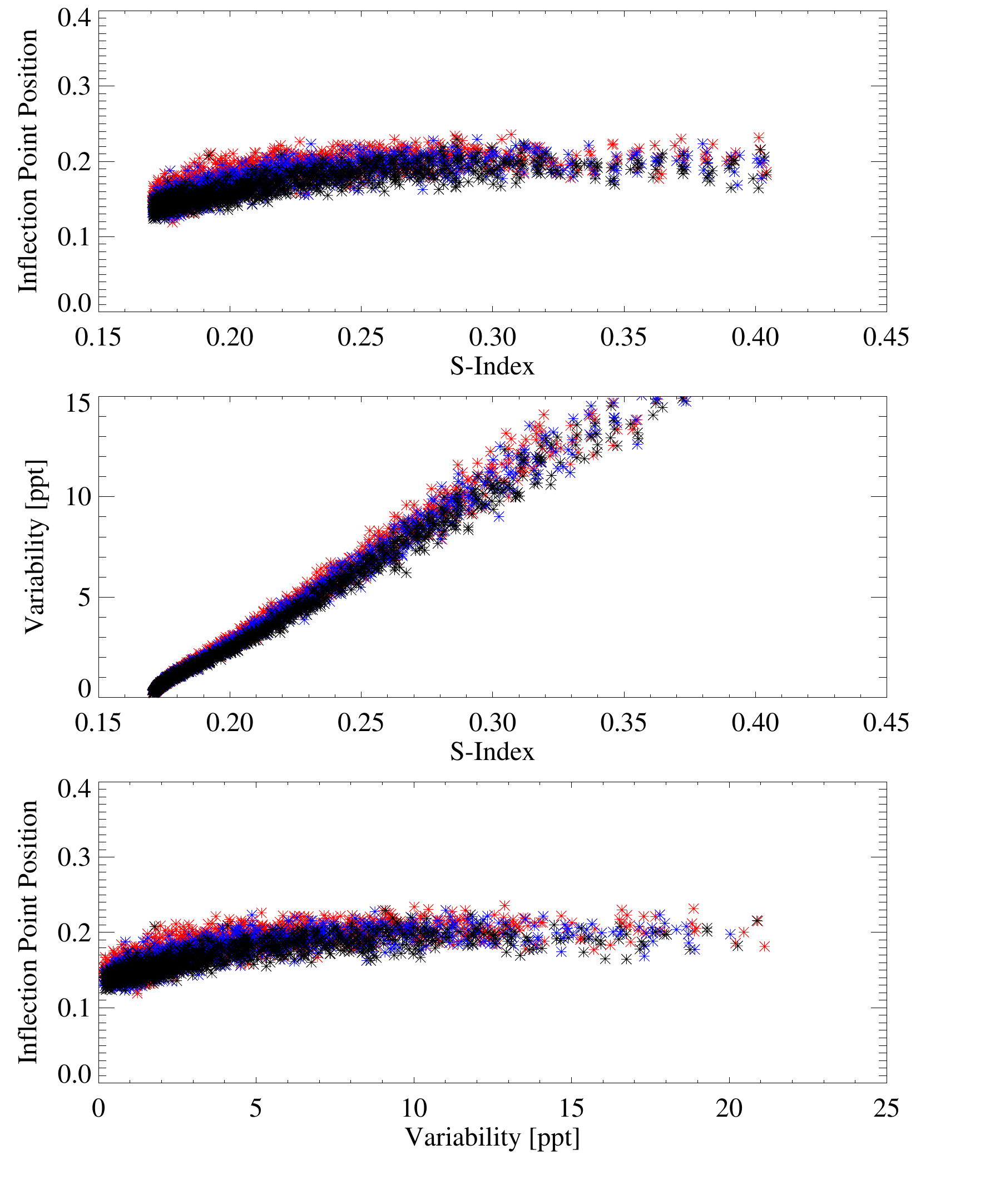}}
\caption{The same dependences as in Fig.~\ref{fig:infl_act} but computed for nine pairs of spot decay rates and ratios between facular and spot lifetimes. For each pair of these parameters,  calculations are performed for five realizations of active regions emergences. Shown are calculations with $T_{\rm fac}/T_{\rm spot}=2$ (black), $T_{\rm fac}/T_{\rm spot}=3$ (blue), and $T_{\rm fac}/T_{\rm spot}=5$ (red). For each $T_{\rm fac}/T_{\rm spot}$ ratio we perform calculations with three values of spot decay rates: 41 MSH/day, 25 MSH/day, and 10 MSH/day.}
\label{fig:infl_act_many}
\end{figure}

Lower and upper panels of Fig.~\ref{fig:infl_act} hint at a seemingly simple way of eliminating the uncertainty in calibration between the stellar rotation period and inflection point position brought about by the unknown facular contribution to stellar variability (see Sect.~\ref{subsect:spot_fac}). One can either estimate the calibration factor from the value of the S-index (if known)  or from the amplitude of the photometric variability. However, all the dependences plotted in Fig.~\ref{fig:infl_act} are produced for a fixed values of the spot decay rate and ratio between facular and spot lifetimes (see above). Both these values are rather uncertain even for the solar case. To take it into account we recalculated all the dependences for a broad range of spot decay rates and ratios between facular and spot lifetimes and plotted them in Fig.~\ref{fig:infl_act_many}. One can see that the resulting dependences are noisier than those plotted in Fig.~\ref{fig:infl_act}. This is because a) the spot decay rate affects the connection between the number of emergences and {\it instantaneous} spot disk-area coverage (which defines the value of the S-index) and b) the ratio between facular and spot lifetimes is in charge of the connection between {\it instantaneous} disk area coverages and those {\it at the time of maximum area}.

All in all, Fig.~\ref{fig:infl_act_many} shows that, despite a significant level of noise, most of the inflection points for stars  {with variability ranges below 3 ppt are located in between 13\% (for faculae-dominated stars) and 21\% (for spot-dominated stars)} of the rotation period. In this respect, we suggest that the best algorithm for determining rotation periods of stars similar to the Sun would be to take a solar value of about 16\%  {(solar value, see Sect.~\ref{subsect:Sun})}, keeping in mind that the intrinsic uncertainty of our method is about 25\%.

We must, however, give some words of caution. We assumed that brightness variations of stars with near solar values of effective temperatures can be calculated by a simple extrapolation of the solar model. In other words, we disregarded the potential presence of active longitudes in the emergence of active regions \citep[we note, however, that the existence of active longitudes have been also proposed for the Sun, see e.g.][]{Sveta_Ilia}, and we assumed a solar distribution of sizes of active regions, solar spot decay rates, as well as solar ratios between facular, spot umbra, and spot penumbra areas. 

The presence of active longitudes might significantly amplify the amplitude of brightness variations and simultaneously make the rotation peak in the power spectra more pronounced. Along the same line, while we do not expect that the size distribution of active regions has a direct impact on the position of the inflection point, it can influence the photometric variability and hence affect the dependence plotted in the lower panel of  Fig.~\ref{fig:infl_act_many}. Finally, there is the critical assumption that the dependence of facular and spot disk-coverages on stellar activity (expressed via the S-index) follow the solar relationships. Any deviations from the assumed  relationships might affect both the position of the inflection point and the amplitude of the photometric variability. We note, however, that solar relationships proved to be very successful for modeling stellar brightness variations on timescales of magnetic activity cycle \citep{Shapiro2014_stars}.




\section{Conclusions}\label{sect:conc}
We have developed a physics-based model for calculating stellar brightness variations. The model is loosely based on the highly successful SATIRE approach for modeling solar brightness variations.

We have utilized our model to show that the rotation signal in the photometric records of stars with near solar fundamental parameters and rotation periods is significantly weakened by a) short lifetimes of spots; b) partial compensation of spot and facular contributions to the rotation signal. Both these factors can also lead to the appearance of ``rogue'' global maxima in the power spectra of stellar brightness variations. These maxima are not associated with the rotation period and can mislead the standard methods for rotation period determination. We construe this as the explanation for the low success rates in detecting rotation periods of stars similar to the Sun \citep{vanSaders2018,Timo2018}.

We have shown that even in the absence of the rotation peak in the power spectra of stellar brightness variations the information about the rotation period is still contained in the high-frequency tail of the power spectrum. In particular, the rotation period can be determined by applying a pre-calculated calibration factor to the frequency corresponding to the inflection point, i.e. the point where the concavity of the power spectrum plotted in the log-log scale changes sign. We have demonstrated that the calibration factor only weakly depends on the parameters describing the evolution of stellar active regions (e.g. their lifetime), stellar disk-area coverage by active regions, and stellar inclination. At the same time the calibration factor depends on the relative areas covered by spots and faculae. On the one hand, this  {introduces intrinsic uncertainty} in the periods determined with our method. On the other hand, the dependence of the calibration factor on he ratio between facular and spot-area coverages allows measuring this ratio in stars with known rotation periods. This might be interesting for constraining the properties of flux emergence in Sun-like stars \citep[see, e.g.][]{Emre2018}. 

 {We have shown that the ratio between  the inflection point position and rotation period is about 0.2--0.23 for the purely spot-dominated stars, which are supposedly much more active than the Sun \citep[see, e.g.][]{Shapiro2014_stars}. The presence of faculae decreases the ratio so that we expect it to be in between 0.13 and 0.21 for stars with near-solar level of photometric variability. Despite a significant uncertainty the main advantage of our method is that it can be used for determining rotational periods of stars with irregular light curves where other available method fails. For such stars we recommend to use solar value of the ratio, i.e. 0.16, which should return rotational period with roughly 25\% uncertainty.}

We intend to further develop the model presented in this study as well as to apply  it to available stellar photometric data. On the theoretical side we plan to a) extend the present study to stars with various fundamental parameters by replacing the \cite{Unruhetal1999} spectra of the quiet Sun and solar magnetic features with recent simulations of stellar spectra \citep[see, e.g.][]{beeck3, norrisetal2017, witzkeetal2018}; b) utilize recent simulations of magnetic flux emergence and transport by \cite{Emre2018} to better describe the evolution of active regions.

On the observational side we plan to a) test our method for the determination of the rotation period against available solar photometric data \citep[see,][]{Eliana1} and stars with known rotation periods; b) apply our method to the sample of Kepler (and, in future, TESS) stars  with unknown rotation periods.

\begin{acknowledgements}
We thank Robert Cameron and Yvonne Unruh for useful discussions. The research leading to this paper has received funding from the European Research Council under the European Union's Horizon 2020 research and innovation program (grant agreement No. 715947). EMAG acknowledges support by the International Max-Planck Research School (IMPRS) for Solar System Science at the University of G{\"o}ttingen. 
It also got a financial support from the BK21 plus program through the National Research Foundation (NRF) funded by the Ministry of Education of Korea. We would like to thank the International Space Science Institute, Bern, for their support of science team 446 and the resulting helpful discussions.
\end{acknowledgements}

\bibliographystyle{aa}

\newpage

\Online

\begin{appendix}

\section{Additional figures}\label{extra}
\begin{figure}
\resizebox{\hsize}{!}{\includegraphics{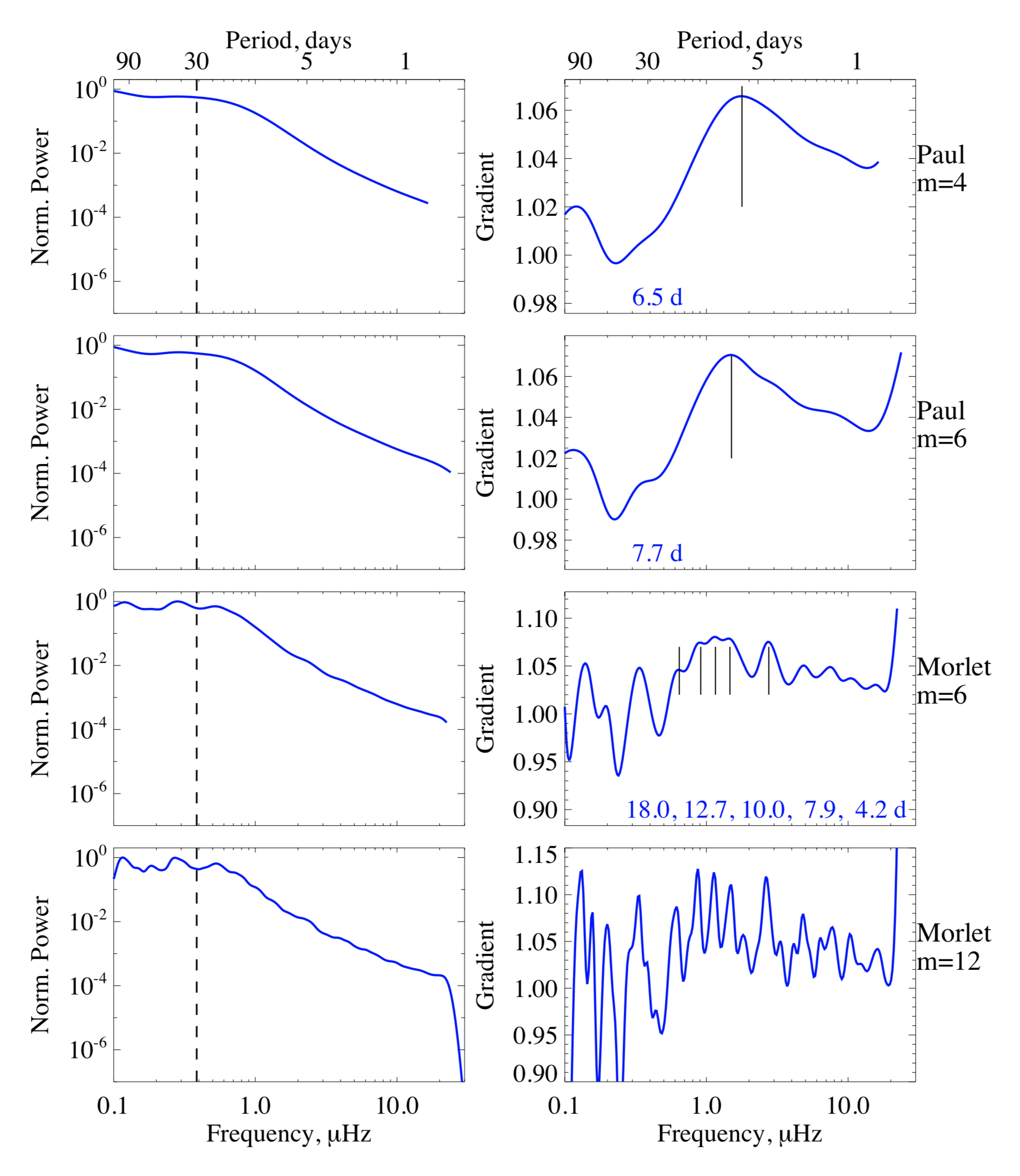}}
\caption{Global wavelet power spectra (left panels) of the light curve from Fig.~\ref{fig:Ex1} and  corresponding  gradients  of  the  power  spectra  (right  panels) calculated with different wavelets. The frequency localization of the utilized wavelet is increasing from the top to bottom panels. As in Fig.~\ref{fig:Ex1} vertical dashed lines in the left panels indicate the rotation period of the modelled star. Vertical solid lines and numbers in the right panels (not shown in the bottom panel) indicate positions of the inflection points.   }
\label{fig:wavelet}
\end{figure}

\begin{figure}
\resizebox{\hsize}{!}{\includegraphics{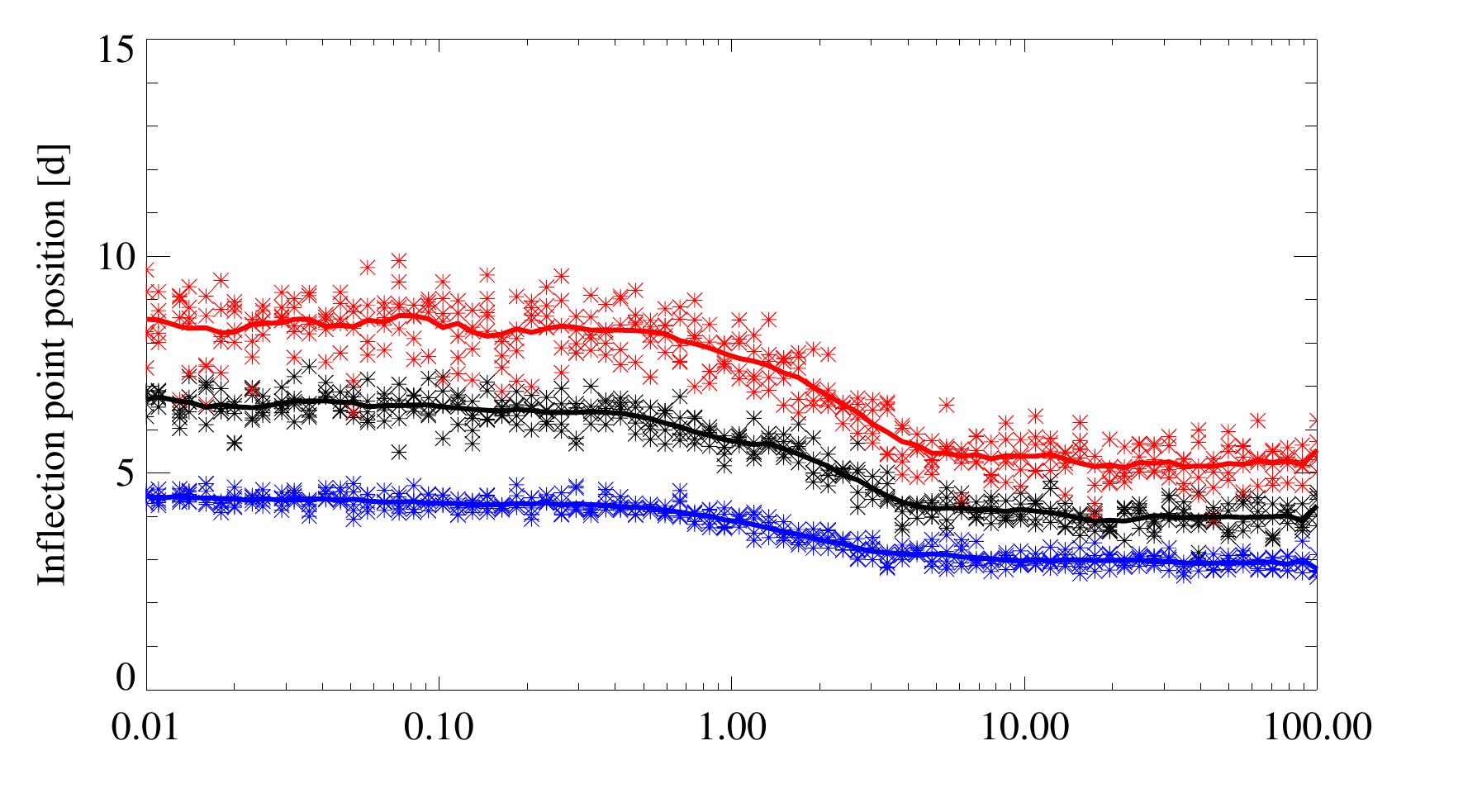}}
\resizebox{\hsize}{!}{\includegraphics{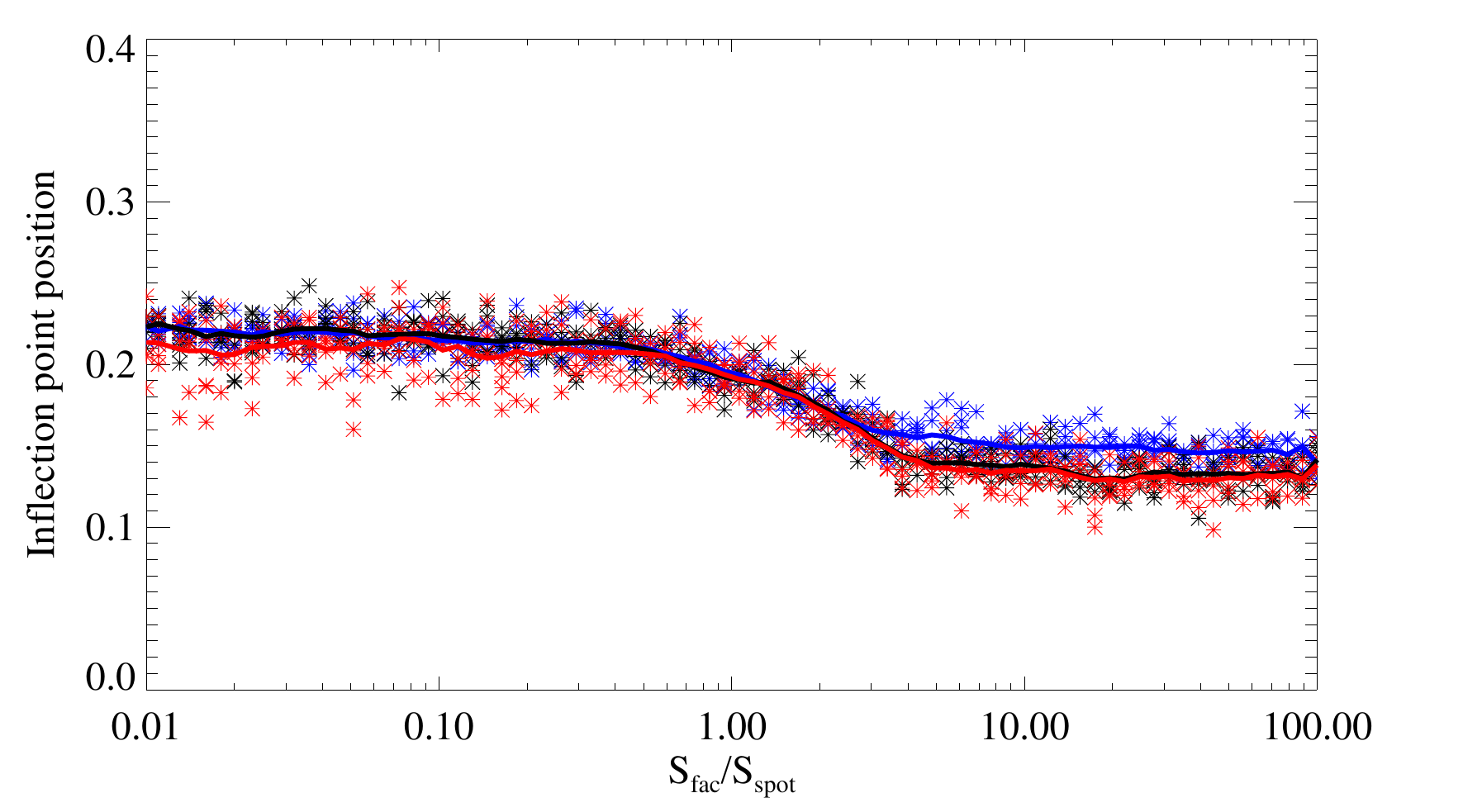}}
\caption{Dependences of the inflection point position on the ratio between facular and spot disk-area coverages at the time of maximum area, $S_{\rm fac}/S_{\rm spot}$, plotted for three values of the rotation period: 20 d (blue), 30 d (black), and 40 days (red). Shown are positions in days (upper panel) and ratios with respect to the rotation period (lower panel). Calculations are performed for a spot decay rate of 25 MSH/day, $T_{\rm fac}/T_{\rm spot}=3$, and mean fractional disk-area spot coverage of 0.3\%. As in Figs.~\ref{fig:fac_spot}--\ref{fig:fac_spot_incl} for each pair of $S_{\rm fac}/S_{\rm spot}$ and rotation periods values, five realization of active regions emergences are shown. Red, black, and blue lines indicate positions of the inflection points averaged over corresponding five realizations.
A small deviation of 20~d curve from 30~d and 40~d curves in the lower panel at high $S_{\rm fac}/S_{\rm spot}$ values can be explained by the insufficient cadence of light curves (4 points per day) for 20~d rotation period and aliasing effect.}
\label{fig:app_period}
\end{figure}

\begin{figure}
\resizebox{\hsize}{!}{\includegraphics{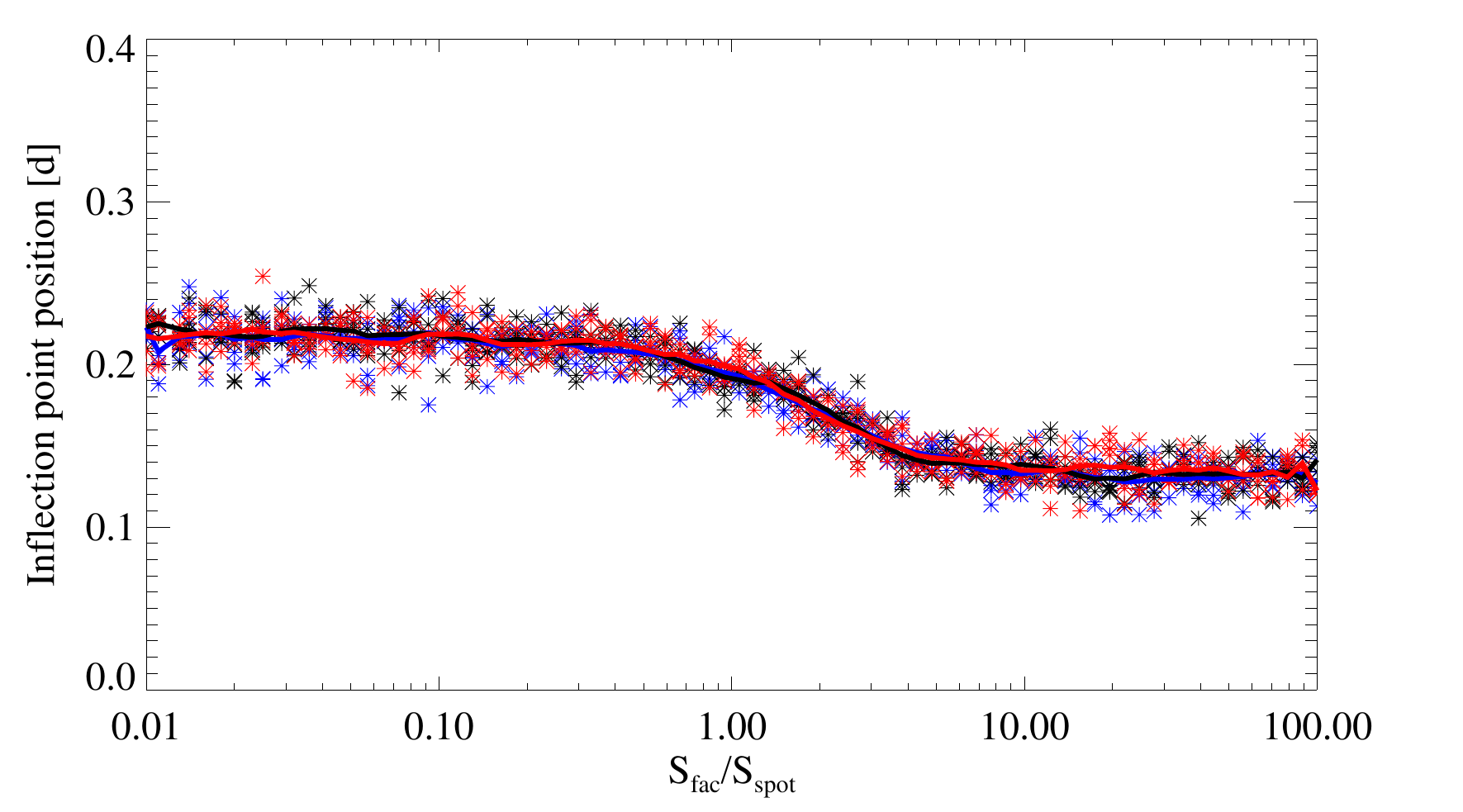}}
\caption{Same as Fig. 8, but for three values of mean fractional disk-area spot coverage: 0.075\% (blue), 0.3\% (black), and 0.75\% (red). Calculations are performed for spot decay rate of 25 MSH/day, $T_{\rm fac}/T_{\rm spot}=3$ and rotation period of 30~d.}
\label{fig:app_coverage}
\end{figure}

\begin{figure}
\resizebox{\hsize}{!}{\includegraphics{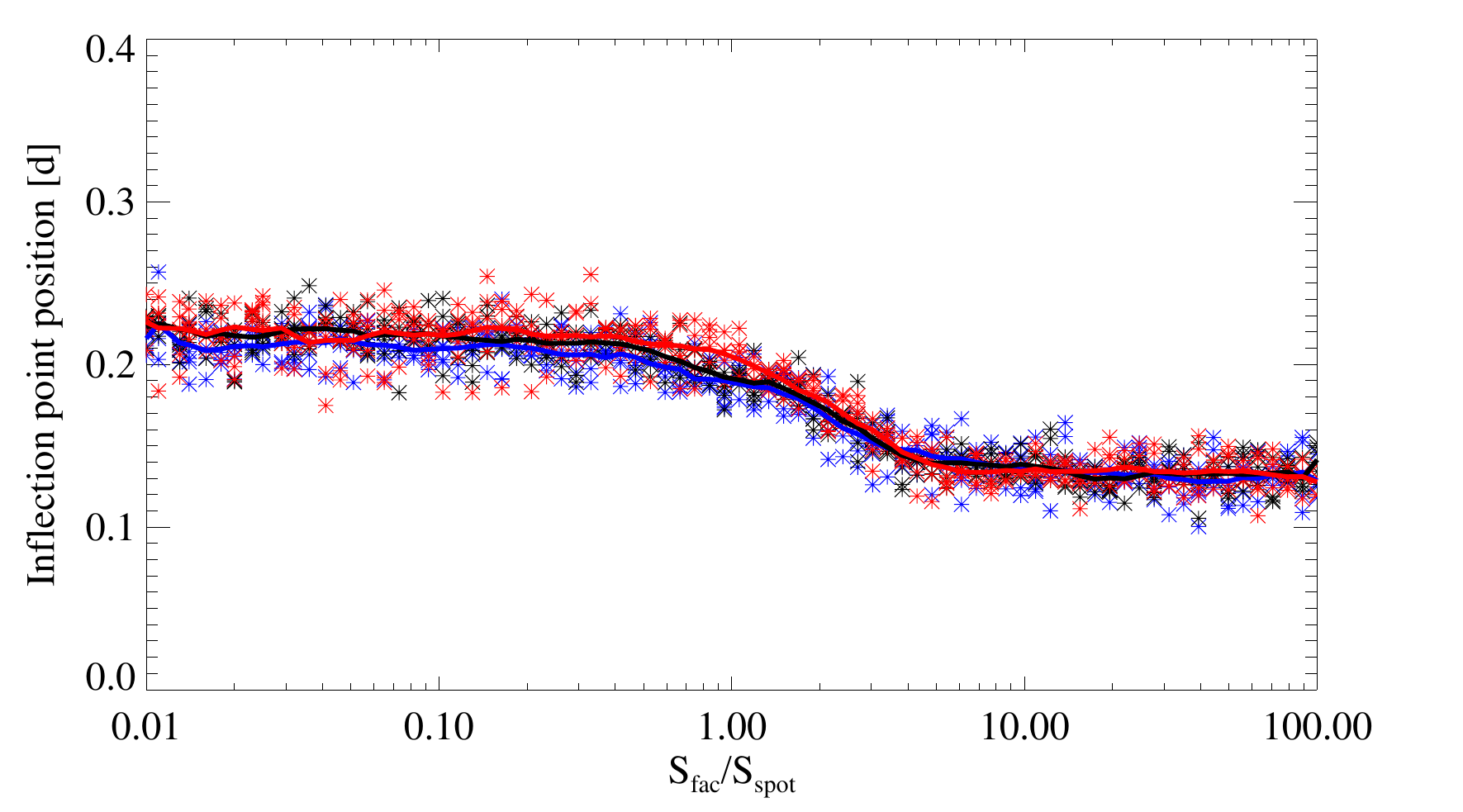}}
\caption{Same as Fig. 8, but for three values of spot decay rate: 10 MSH/day (blue), 25 MSH/day (black), 41 MSH/day (red). Calculations are performed for mean fractional disk-area spot coverage of 0.3\%, $T_{\rm fac}/T_{\rm spot}=3$ and rotation period of 30~d.}
\label{fig:app_decay}
\end{figure}

\begin{figure}
\resizebox{\hsize}{!}{\includegraphics{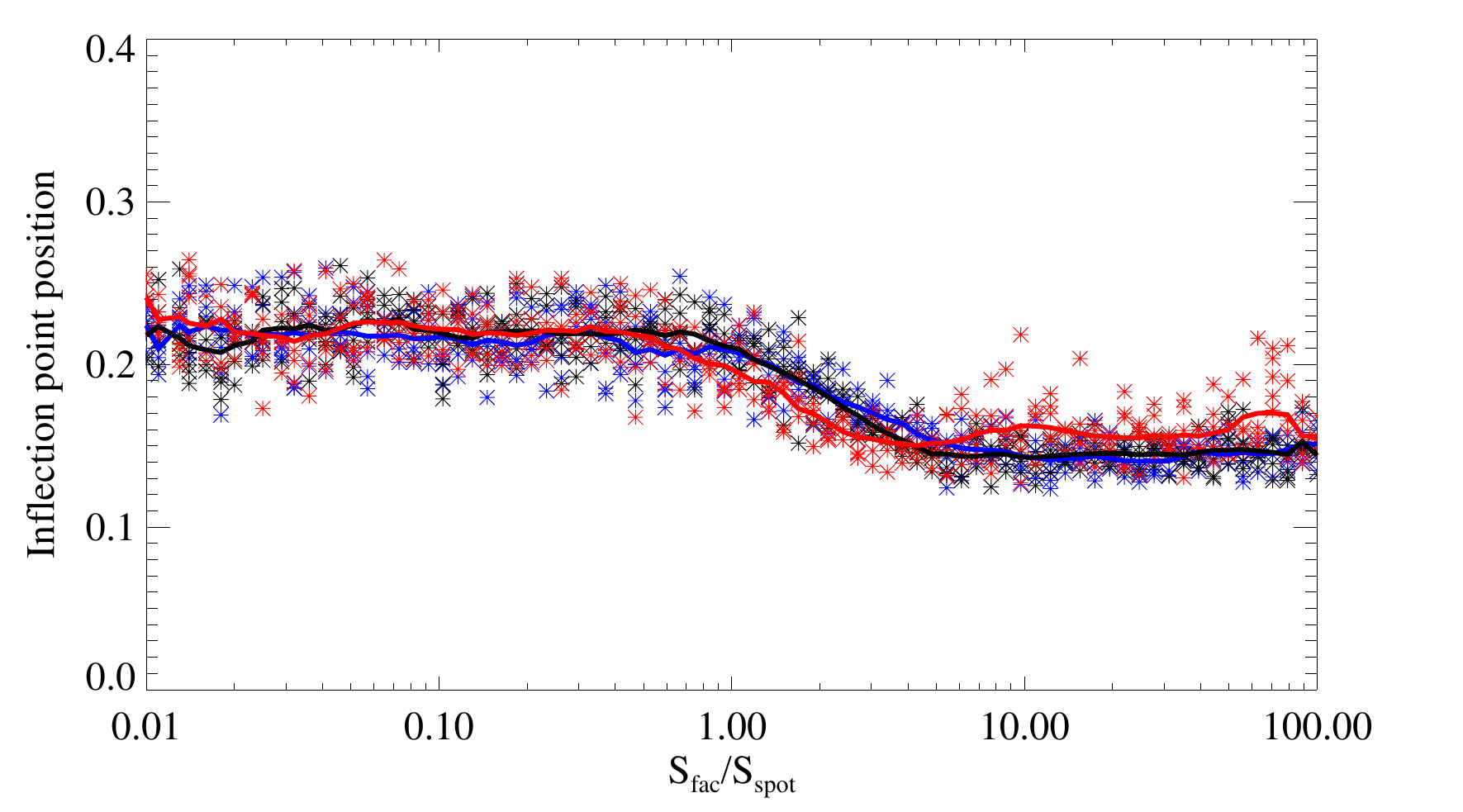}}
\caption{Same as Fig. 8, but calculated assuming that all emergences of active regions happen at the equator (blue), latitudes $\pm 30^{\circ}$ (black), and latitudes $\pm 60^{\circ}$ (red). Calculations are performed for mean fractional disk-area spot coverage of 0.3\%, $T_{\rm fac}/T_{\rm spot}=3$ and rotation period of 30~d.}
\label{fig:app_lat}
\end{figure}

The section includes Figs.~\ref{fig:wavelet}--\ref{fig:app_lat}.

\section{Solar values of facular to spot area ratio at the time of maximum area}\label{sect:fac_to_spot}
As discussed in Sect.~\ref{subsect:fac_tr} the directly observed {\it instantaneous} ratio between solar disk coverages by faculae and spots is different from the ratio between facular and spot areas in individual magnetic regions {\it at the time of their maximum area}, $S_{\rm fac}/S_{\rm spot}$. At the same time in Sect.~\ref{sect:rel}  we have shown  that the position of the inflection point is defined by the ratio {\it at the time of maximum area}, $S_{\rm fac}/S_{\rm spot}$ and thus we need to know solar value of $S_{\rm fac}/S_{\rm spot}$ to check whether position of the inflection point in the power spectrum of solar brightness variations  is consistent with the model presented here. In this section we show how to determine solar value of $S_{\rm fac}/S_{\rm spot}$ from the observed {\it instantaneous} records of solar disk coverages by spots and faculae (see Sect.~\ref{subsect:Sun}). 

First we employ the model setup described in Sect.~\ref{subsect:fac_tr} to calculate the power spectra of modeled facular and spot disk area coverages as they would be seen along the stellar rotation axis and from the stellar equatorial plane. We consider $P_{\rm rot}=30$~d, $S_{\rm fac}/S_{\rm spot}=3.4$, $T_{\rm fac}/T_{\rm spot}=3$ case, adopt log-normal distribution of spot sizes from \cite{BaumannSolanki2005}  and 25 MSH/day for the sunspot decay rate.  

\begin{figure*}
\resizebox{\hsize}{!}{\includegraphics{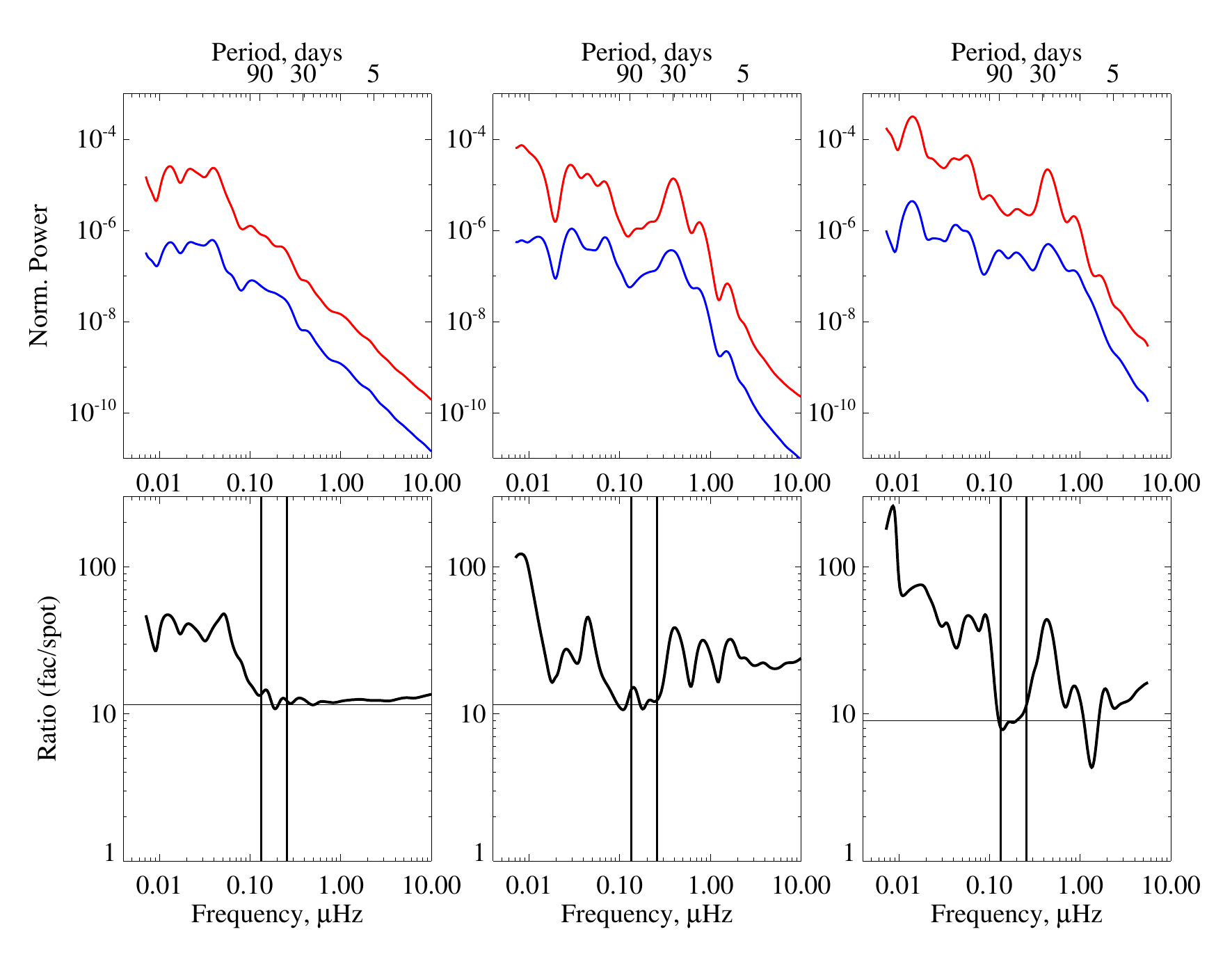}}
\caption{Power spectra of facular (red) and spot (blue) disk area coverages for a modeled star observed along its rotational axis (left upper panel), a modeled star observed from its equatorial plane (middle upper panel), and for the Sun deduced from the  SDO/HMI observations (right upper panel). The corresponding ratios between the power spectra of facular and spot  disk area coverages are plotted in the lower panels. The vertical black lines in the lower panel denote the interval between 45 and 90 days. The horizontal black line in the left and middle lower panel denotes the $S_{\rm fac}^2/S_{\rm spot}^2$ value used in the simulations (see text for more details). The horizontal black line in the right lower panels denotes ratio value of 9.}
\label{fig:spot_fac}
\end{figure*}

Left panels of Fig.~\ref{fig:spot_fac} show the global wavelet (Morlet, 6th order) power spectra (top) of {\it instantaneous} disk coverages by spots and faculae observed along rotation axis of a modeled star (so that the rotational modulation does not affect the power spectra) as well as their ratio (bottom). One can see that the ratio is roughly constant and is equal to $S_{\rm fac}^2/S_{\rm spot}^2$ up to the period of about 90 days (i.e. $3 P_{\rm rot}$). This is not surprising since the decay of magnetic features only affects the power spectrum at timescales larger than the decay time. 

More strictly, when observing along the rotation axis, the disk area coverage by the individual magnetic feature is proportional to the product of the unit step function (i.e. function which returns 0 before the emergence of the magnetic feature and 1 after the emergence) and a function describing linear decay. The power spectral density of such a right-triangle function (hereafter function 
${\cal F}_1(t)$) is proportional to:
\begin{equation}
{\cal D}(\nu) \sim   \frac{S_{\rm feature}^2}{x^2} + \frac{S_{\rm feature}^2}{x^4} \left (  \sin(x/2)^2 +  \sin((x-\pi/2)/2)^2       \right ),
\label{eq:pol}
\end{equation}
where $S_{\rm feature}$ is the disk area coverage of the feature {\it at the time of maximum area}, $x \equiv  2 \pi \, T_{\rm dec} \nu$, and $T_{\rm dec}$ is a decay time of magnetic feature (see Eq.~\ref{eq:lin}). 

At $x>>1$ (which corresponds to $P<<2 \pi \, T_{\rm dec}$, where $P \equiv 1/\nu$) the second term on the right-hand side of Eq.~\ref{eq:pol} becomes negligibly small in comparison to the first term. Consequently,  the corresponding power spectral density ${\cal D}(\nu)$  drops with frequency as $ 1/\nu^2$ independently of the decay time of magnetic feature. Hence, the ratio between power spectra of facular and spot disk area coverages of an active region (consisting of facular and spot features, see Sect.~\ref{sect:fac}) is equal to $S_{\rm fac}^2/S_{\rm spot}^2$. We note that the  power spectra of the observed {\it instantaneous} facular and spot disk coverages are brought about by the superposition of the contributions from many incoherently emerging active regions. Therefore the ratio of high-frequency parts of the power spectra represents the mean  $ S_{\rm fac}/S_{\rm spot} $ value over all active regions. Since the contribution of magnetic features to the disk coverage is proportional to their size this mean value is weighted towards larger active regions.

Let us now consider the case of the observations from the stellar equatorial plane (i.e. the solar case since the $\approx 7.25^{\circ}$ angle between the solar equator and ecliptic can be neglected in our analysis). Middle panels of Fig.~\ref{fig:spot_fac} show the global wavelet power spectra of facular and spot disk coverages resulting from the same realization of active regions emergences as shown in the left panels, but now the active regions are observed from the equatorial plane of the modeled star. One can see that the ratio between facular and spot power spectra is strongly affected by rotation at 
periods below 45 d (i.e. 3/2 $P_{\rm rot}$) but is basically not affected by the rotation at periods between 45 d and 90 d. Below we give an explanation of such a behavior.

The disk area coverage by a single transiting magnetic feature for a star observed from the equatorial plane is  proportional to the product of three functions: a. the right-triangle function ${\cal F}_1(t)$ (with power spectral density given by Eq.~\ref{eq:pol}); b. a function which returns zero during half of the period when the feature is at the far-side of the star and 1 during another half of the period when the feature is on the near-side of the star (i.e. shifted by 0.5 square wave function, hereafter function ${\cal F}_2(t)$); c. function describing foreshortening effect (hereafter function ${\cal F}_3(t)$).

The Fourier transform of the  square wave function contains only odd-integer harmonics of the form 
$\pm (2k-1) \nu_{\rm rot} $ where $\nu_{\rm rot}=1/P_{\rm rot} $. The shift by 0.5 brings about an additional zero frequency component so that the Fourier transform of function ${\cal F}_2(t)$ contains 
zero frequency and odd-integer harmonics. The foreshortening function ${\cal F}_3(t)$ is proportional to the cosine of the angle between the direction from the centre of the star to the observer and to the magnetic feature and contains only $\pm \nu_{\rm rot} $ components. Multiplication in the time domain corresponds to the convolution in the frequency domain. Consequently, the Fourier transform of the product of  ${\cal F}_2(t)$ and  ${\cal F}_3(t)$ functions contains all harmonics of the rotational period $\pm k  \nu_{\rm rot}$ and zero frequency component.

All in all, the disk area coverages observed from the stellar equatorial plane can be obtained by multiplying disk area coverages observed along the stellar rotation axis (which are proportional to ${\cal F}_1(t)$) with a function containing zero frequency component and harmonics of the rotational frequency. As discussed above, the power spectra of facular and spot disk coverages are proportional to each other with the exception of the $[0, \nu_{\rm rot}/3]$ interval. After the convolution the signal in this interval will be propagated to $[\mid \pm k \, \nu_{\rm rot} \mid , \mid  \nu_{\rm rot}/3 \pm \nu_{\rm rot} \mid ]$ (i.e. $[0, \nu_{\rm rot}/3]$, $[\nu_{\rm rot}, 4/3 \, \nu_{\rm rot}]$, $[2 \, \nu_{\rm rot}, 7/3 \, \nu_{\rm rot}]$..., and $[2/3 \, \nu_{\rm rot}, \nu_{\rm rot}]$, $[5/3 \, \nu_{\rm rot}, 2 \, \nu_{\rm rot}]$, ...) intervals. Interestingly, the signal does not propagate to the  $[1/3 \, \nu_{\rm rot}, 2/3 \, \nu_{\rm rot}]$ (or  $[3/2 \, P_{\rm rot}, 3  \, P_{\rm rot}]$) interval. This explains the curious behavior of the ratio between power spectral density of  facular and spot disk coverages in this interval shown in the middle lower panel of Fig.~\ref{fig:spot_fac}: 
neither the decay of magnetic features nor the stellar rotation affects it.

The power spectra and their ratio shown in right panels of Fig.~\ref{fig:spot_fac} are calculated using solar disk area coverages obtained by  \cite{yeoetal2014} using solar magnetograms and continuum images recorded by the Helioseismic and Magnetic Imager onboard the Solar Dynamics Observatory (SDO/HMI) for the period from May 2010 till August 2014. In agreement with the previous discussion the ratio between the two power spectra is roughly constant in the time interval between 45 d and 90 d and corresponds to $S_{\rm fac}/S_{\rm spot}$ value of 3.

\begin{figure}
\resizebox{0.9\hsize}{!}{\includegraphics{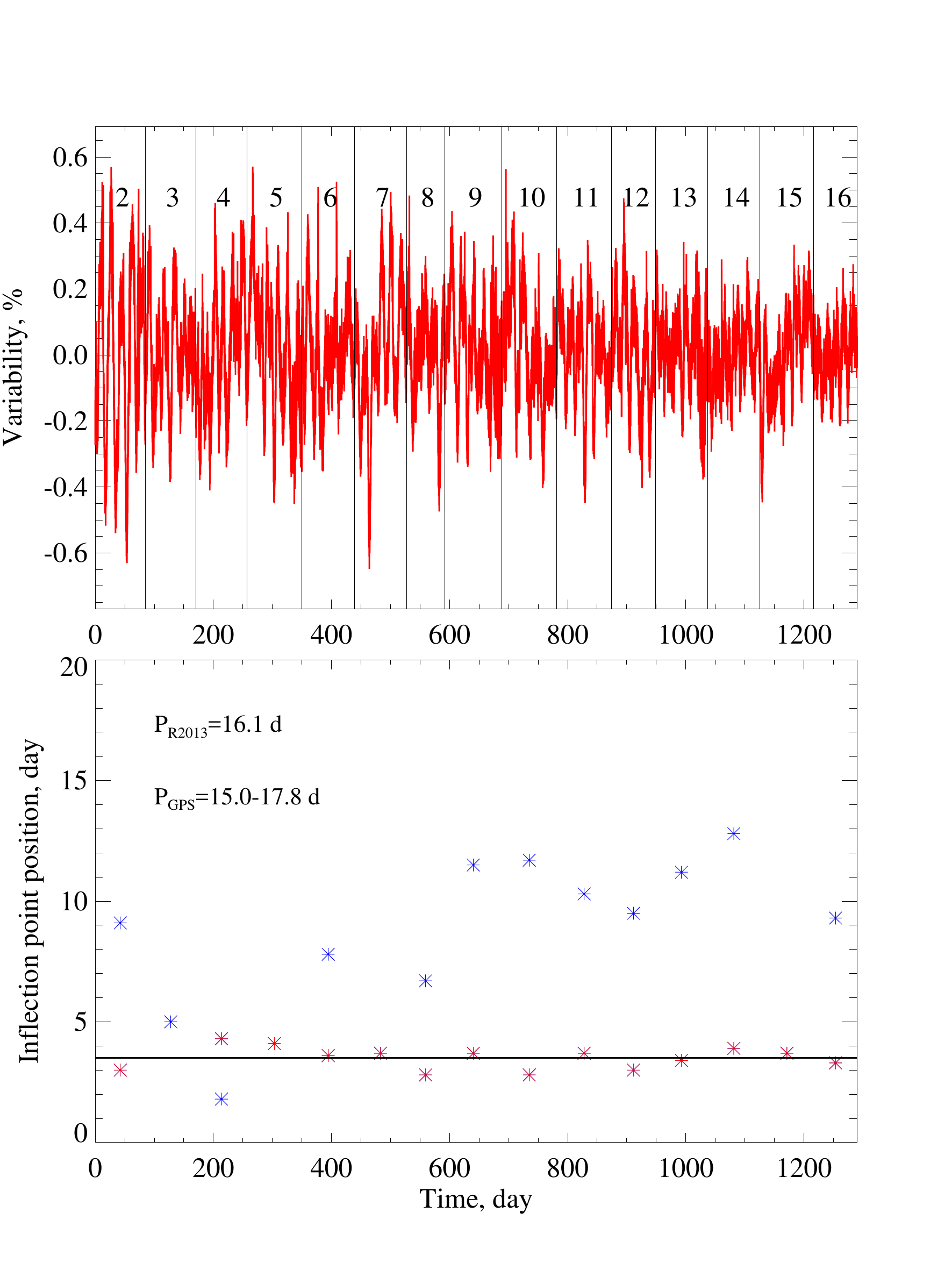}}
\caption{The light curve of KIC2141852 (upper panel) and positions of the inflection points for each of the Kepler quarters (lower panel). The Kepler quarters 2--16 are numbered in the upper panel and separated  by the vertical black lines. The asterisks in the lower panel correspond to the positions of inflection points. The value of the outlier-resistant mean of the inflection point positions in the low-period branch (see text for more details) is indicated with the horizontal black line. Red asterisks correspond to the inflection points utilized for calculating the outlier-resistant mean value, blue asterisks are trimmed as outliers. The rotation period range returned by our method ($P_{\rm GPS}$) and period from \cite{Timo2013} ($P_{\rm R2013}$) are listed in the lower panel. }
\label{fig:ex1}
\end{figure}

\begin{figure}
\resizebox{0.9\hsize}{!}{\includegraphics{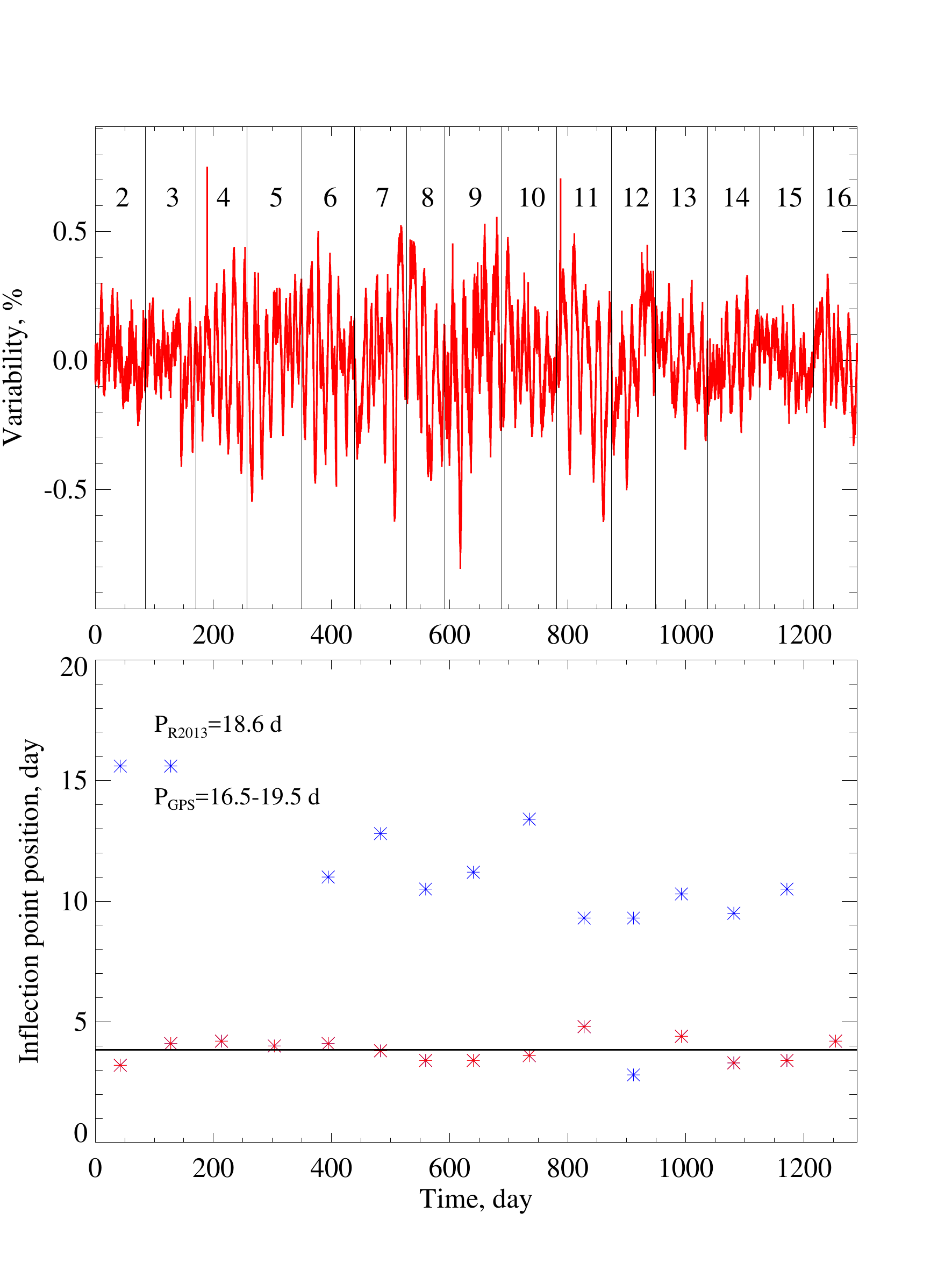}}
\caption{The same as \ref{fig:ex1} but for KIC2553816.}
\label{fig:ex2}
\end{figure}

\begin{figure}
\resizebox{0.9\hsize}{!}{\includegraphics{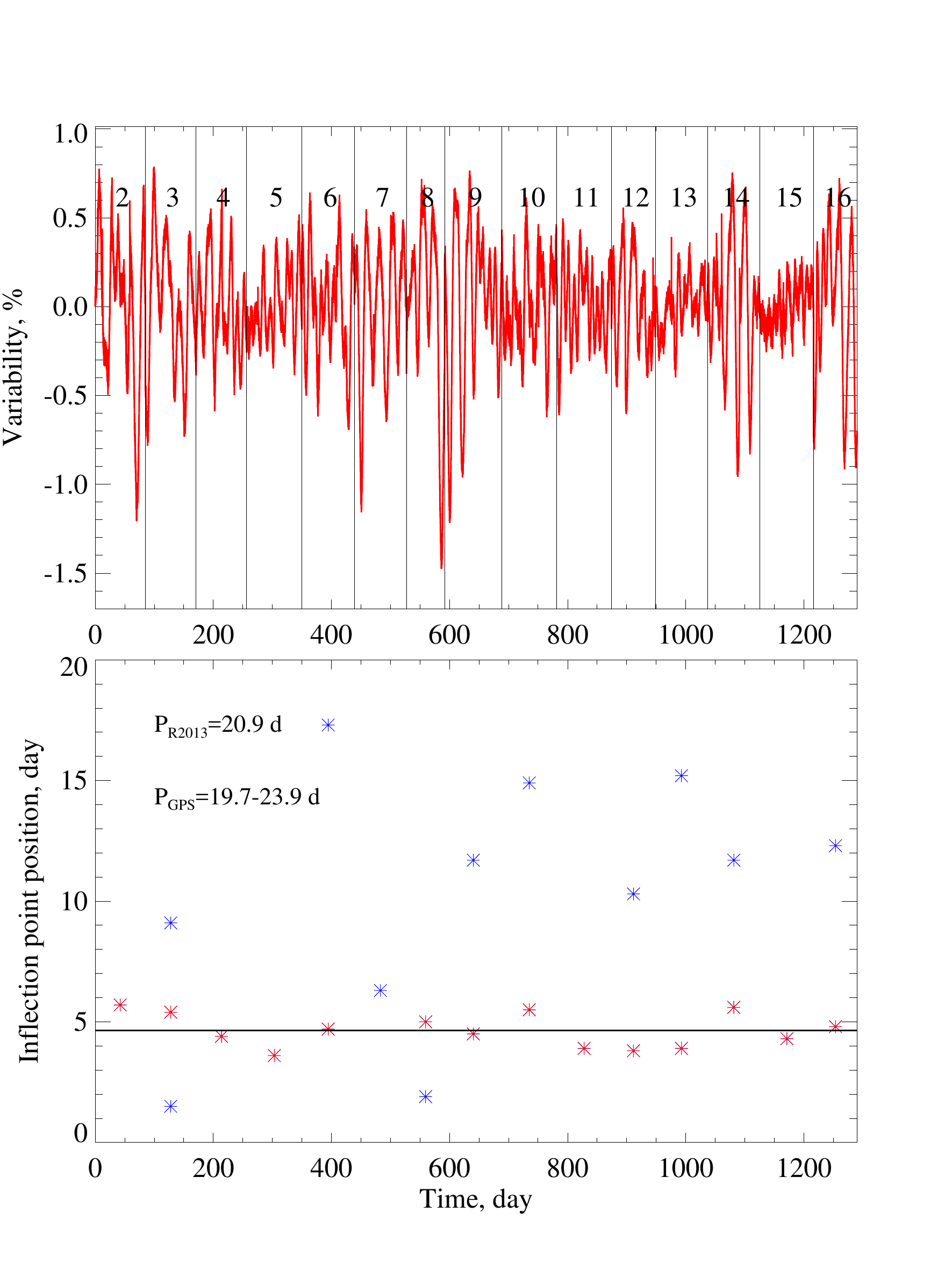}}
\caption{The same as \ref{fig:ex1} but for KIC2992964.}
\label{fig:ex3}
\end{figure}

 {
\section{Three examples of the GPS method application to the Kepler stars}\label{Examples}
An extensive test of our method for determining stellar rotation periods will be in focus
of the forthcoming publications. In particular, we will analyze dependence of the inflection point position on the photometric variability and test the dependence established in Sect.~5.2. Here we, as an example, apply our method to stars significantly more variable than the Sun with presumably spot-dominated variability.

In the upper panels of Figs.~\ref{fig:ex1}--\ref{fig:ex3} we show the Kepler light curves of 
KIC2141852 (Fig.~\ref{fig:ex1}), KIC2553816 (Fig.~\ref{fig:ex2}), and KIC2992964 (Fig.~\ref{fig:ex3}). The data for 15 Kepler quarters (quarter 2 -- quarter 16) have been downloaded from the MAST portal\footnote{https://mast.stsci.edu/portal/Mashup/Clients/Mast/Portal.html} and  reduced with the PDC-MAP pipeline \citep{PDC-MAP1,PDC-MAP2}. The lower panels of  Figs.~\ref{fig:ex1}--\ref{fig:ex3} show the position of the inflection points for each of the Kepler quarters. The behavior of the inflection points is very similar to that shown for the synthesized light curves in Fig.~\ref{fig:example_fac}. Namely, inflection points fluctuate around the mean position, and from time to time ``rogue'' inflection points, mainly corresponding to the high-period branch, appear. We note that a large number of high-period inflection points imply that the lifetime of magnetic features on considered stars is comparable or larger than their rotational periods (see Sect.~\ref{subsect:decay}). This is consistent with highly regular light curves of considered stars. 

We have calculated the outlier-resistant mean of the low-period  inflection point positions as well as the  {standard} error of the mean. Since we expect that the variability of our exemplary stars is spot-dominated, the ratio between the inflection point position and the rotational period should lie between 0.2 and 0.23 (see Fig.~\ref{fig:fac_spot}). We applied these factors to the mean position of the inflection point taking the  {standard} error of the mean into account. The resulting ranges for the rotation periods ($P_{\rm GPS}$) are indicated in  Figs.~\ref{fig:ex1}--\ref{fig:ex3} and compared to the periods ($P_{\rm R2013}$) reported by \cite{Timo2013}. One can see that periods from \cite{Timo2013} are  within the range given by our method for all three considered stars.}

\end{appendix}

\end{document}